\documentclass[superscriptaddress,twocolumn,aps,prd,linenumbers,nofootinbib]{revtex4}	

\usepackage{graphics}
\usepackage[caption=false]{subfig}
\usepackage{float}
\usepackage{amsmath}
\usepackage{amssymb}
\usepackage[normalem]{ulem} 
\usepackage[mathscr]{euscript}
\usepackage{acronym}
\usepackage[dvipsnames]{xcolor}
\floatstyle{boxed} 
\usepackage{lipsum}
\usepackage[section]{placeins}
\usepackage{mathtools}
\usepackage{multirow}
\usepackage{graphicx}
\usepackage{xspace}
\usepackage{tikz}
\usepackage{rotating}
\usepackage{microtype}
\usetikzlibrary{shapes,arrows}
\usetikzlibrary{matrix}
\usetikzlibrary{er,positioning}
\DeclareMathAlphabet\mathpazo{OML}{zplm}{m}{it}
\def\thercsid{\relax}
\def\rcsid#1{\def\next##1#1{\def\thercsid{##1}}\next}
\rcsid$Id: QNM.tex,v  Doc No. $
\renewcommand{\today}{\number\day\space\ifcase\month\or
January\or February\or March\or April\or May\or June\or
July\or August\or September\or October\or November\or December\fi
\space\number\year}	

\usepackage[colorlinks, pdfborder={0 0 0}, plainpages=false]{hyperref}
\usepackage{graphicx}
\usepackage[caption=false]{subfig}
\usepackage{xspace}

\newcommand{\Mark}[1]{\textcolor{Cerulean}{#1}}
\newcommand{\saul}[1]{\textcolor{red}{#1}}

\usepackage{mathrsfs}

\newcommand{\beq}{\begin{equation}}
\newcommand{\eeq}{\end{equation}}
\newcommand{\bit}{\begin{itemize}}
\newcommand{\eit}{\end{itemize}}

\newcommand{\Ep}{\mathcal E}

\newcommand{\fni}{\mathscr{I^+}}

\newcommand{\FlowChartFigure}{%
\begin{figure}[!htbp]\vspace*{1\baselineskip}
\centering
\begin{tikzpicture}[->,>=stealth',shorten >=1pt,auto,node distance=2cm]
 \node[entity,fill=violet!5] at (0,0) (A) {Generic Spacetime};
  \node[entity,fill=violet!11.25] at (3,-1.5) (B) {Algebraically Special};
 \node[entity,fill=violet!17.5] at (0,-4) (C) {Petrov Type D};
 \node[entity,fill=violet!23.75] at (3,-5.5) (D)  {Kerr-NUT};
  \node[entity,fill=violet!30] at (0,-7) (E)  {Kerr};
 
\draw [line width=0.2mm] (A) -- (B) node [midway, fill=white] {Speciality Index Eq.~\eqref{eq:Speciality Index} $N_d = 1$};
\draw [line width=0.2mm](B) -- (C) node [near start, fill=white,text width=5cm] {Type D 1 Eq.~\eqref{eq:typed1} \;  $N_d = 1$ Type D 2 Eq.~\eqref{eq:typed2}  $N_d = 2$ Type D 3 Eq.~\eqref{eq:typed3}  $N_d = 3$ Type D 4 Eq.~\eqref{eq:typed4}  $N_d = 4$ };
\draw [line width=0.2mm](C) -- (D) node [midway, fill=white,text width=5cm] {Kerr 1 Eq.~\eqref{eq:Kerr1}  $N_d = 2$};
\draw [line width=0.2mm](D) -- (E) node [near start, fill=white,text width=5cm] {Kerr 2 Eq.~\eqref{eq:Kerr2}  $N_d = 2$ Kerr 3 Eq.~\eqref{eq:Kerr3}  $N_d = 2$};                               
\end{tikzpicture}
  \caption{ The set of conditions for a slice to be locally isometric to Kerr. The nodes refer to the resulting type of spacetime when the conditions on each edge, given by their name and equation in the text, are met. For example, a spacetime must meet all four of the conditions specified in the edge from Algebraically Special to Petrov Type D to belong to the type D subset of algebraically special spacetimes. In numerical applications, the failure of these Kerrness conditions to be met gives a set of respective Kerrness measures, where larger measures denote greater deviation from Kerr. For each measure, we give $N_d$, the number of numerical derivatives beyond the first derivatives of the metric needed to evaluate it, which corresponds to the numerical noise level in the measure, with higher derivative powers giving more numerical noise.}
  \label{fig:FlowChart}
\end{figure}
}
\newcommand{\alg}{%
\begin{figure}[!htbp]
\subfloat{\includegraphics[width=0.5\textwidth]{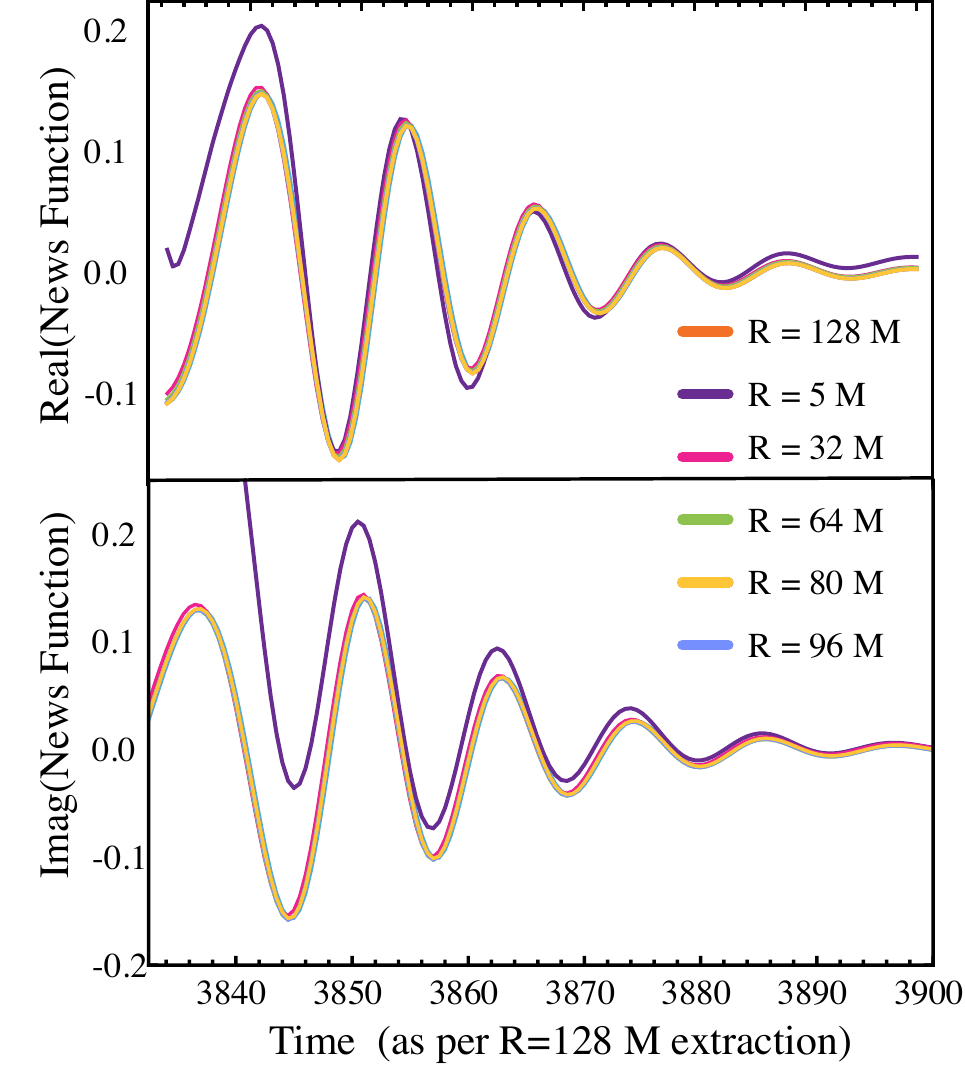} } 
\caption{The $l=m=2$ mode of the news function seen at $\fni$ extracted from worldtube boundaries of $R = 5\,M$,~$32\,M$,~$64\,M$,~$80\,M$, ~$96\,M$~and~$128\,M$.  The horizontal axis corresponds to the time stamps associated with the news function corresponding to CCE from $R = 128\,M$. The \textbf{top panel} shows the real part and the \textbf{bottom panel} shows the imaginary part of the news function. The alignment of news functions has been done such that the overlap is maximized. The transformation that changes the gauge from a non-inertial to an inertial observer has not been applied to any of the extractions. All of the extractions beginning with $R = 32\,M$ seem to agree with one another. Notice that the amplitude of the news function extracted from $R = 5\,M$ deviates from the other extractions, especially in the first cycle. Nevertheless, the phase evolution between the news function from extraction radii seem to agree.  }
\label{fig:AlignmentAtMaximumOftheNews}
\end{figure}
}

\newcommand{\spread}{%
\begin{figure}[!htbp]
\subfloat{\includegraphics[width=\columnwidth, keepaspectratio]{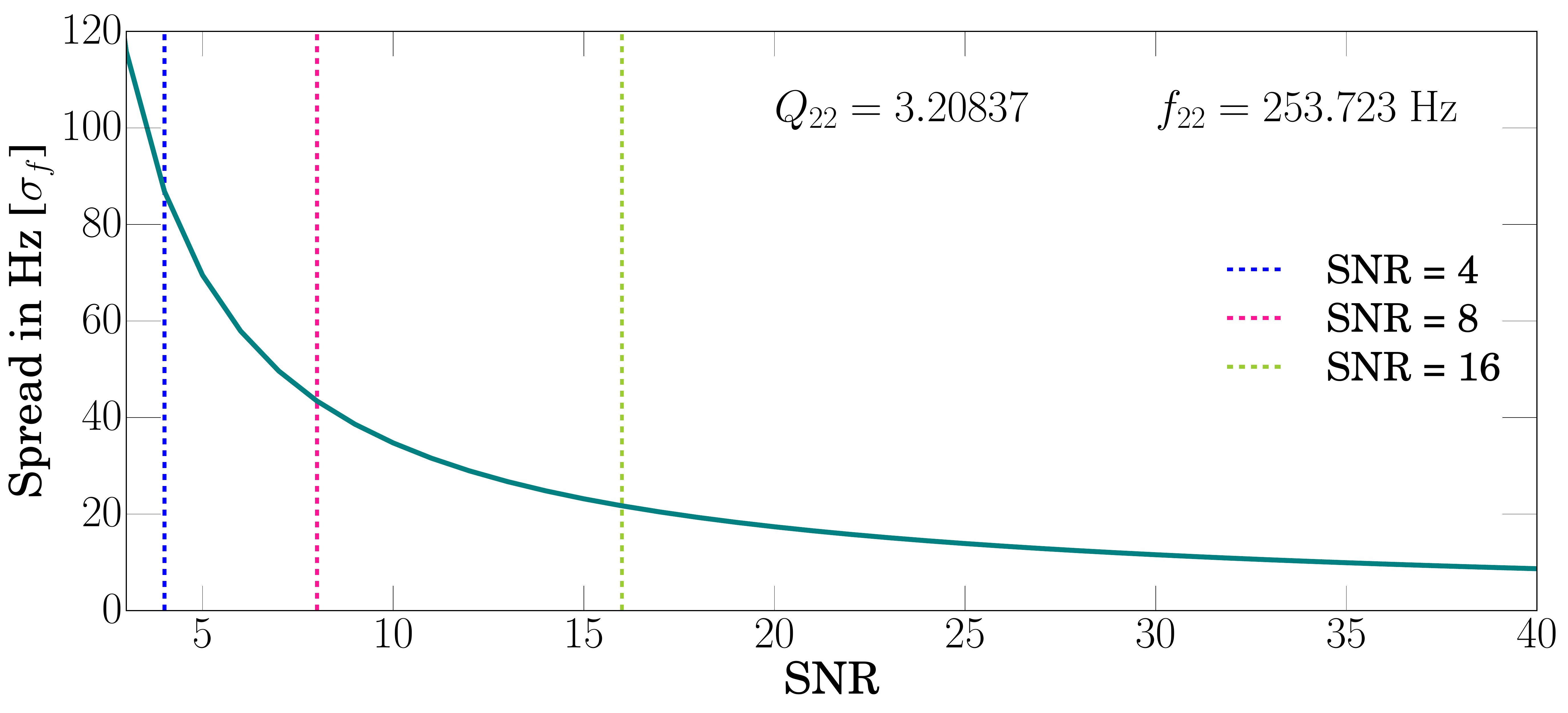}}
\caption{Spread in estimation of dominant mode frequency as a function of SNR. We present the spread, $\sigma_{f}$ in the estimation of frequency calculated using Fisher information matrix formalism. We should the increase in spread with decreasing SNR, providing the rough intuition on the implication of Fig.~\ref{fig:PercentageSNR} on parameter estimation.}
\label{fig:spread}
\end{figure}}

\newcommand{\postItSpecial}{%
\begin{figure*}[!htbp]
\subfloat{\includegraphics[width=\columnwidth,keepaspectratio]{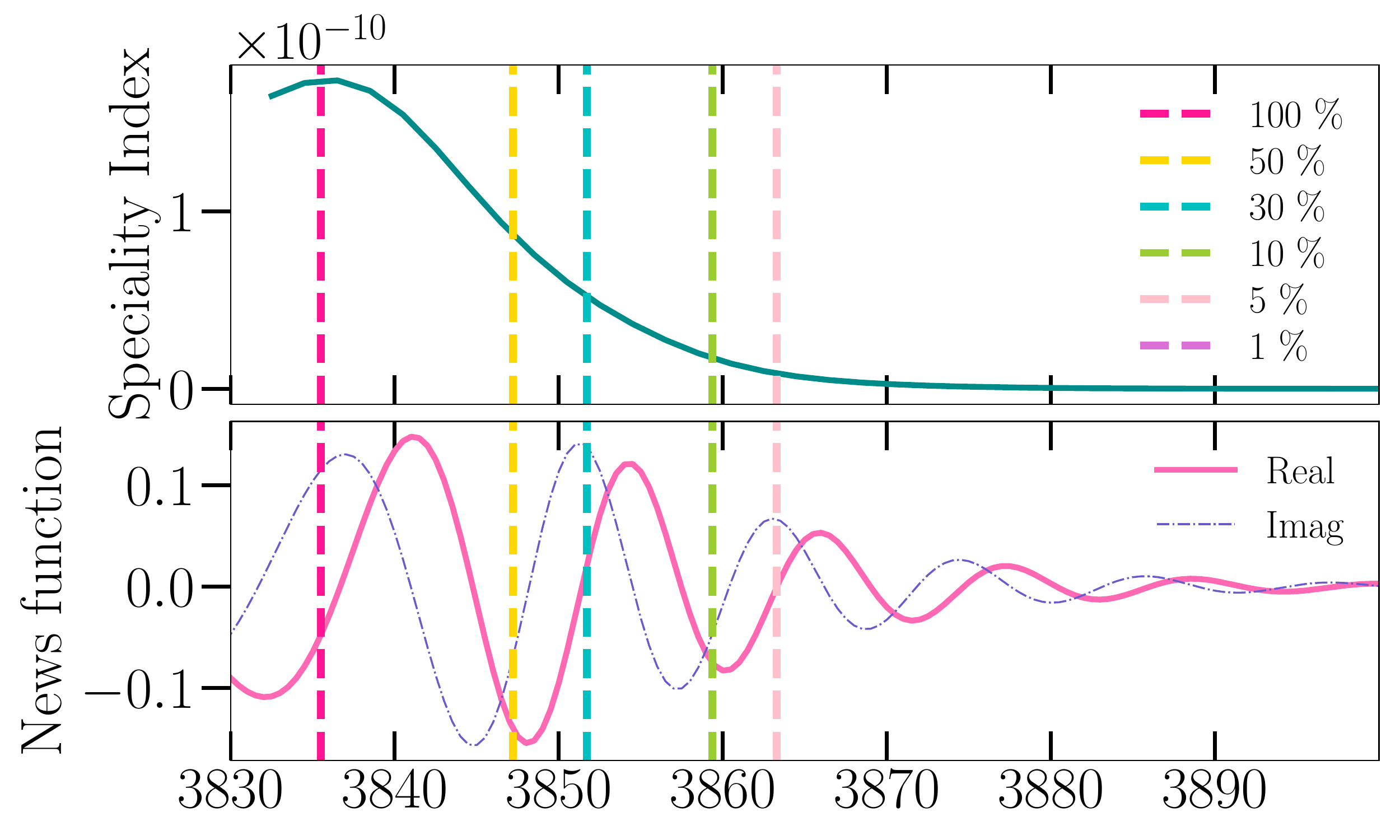} }
\caption{This figure is similar to Fig.~\ref{fig:PostItPanel} but for Specialty Index. We plot this separately as it is an independent measure and decays rapidly compared to the other measures. Further, we do not indicate the $1\%$ of peak line because of numerical noise (cf. Fig.~\ref{fig:Rainbow}) which leads to unreliable root finding for time of percentage decrease.}
\label{fig:specialityIndex}
\end{figure*}}

\newcommand{\perOnWave}{%
\begin{figure*}
\subfloat{\includegraphics[width=\textwidth,keepaspectratio]{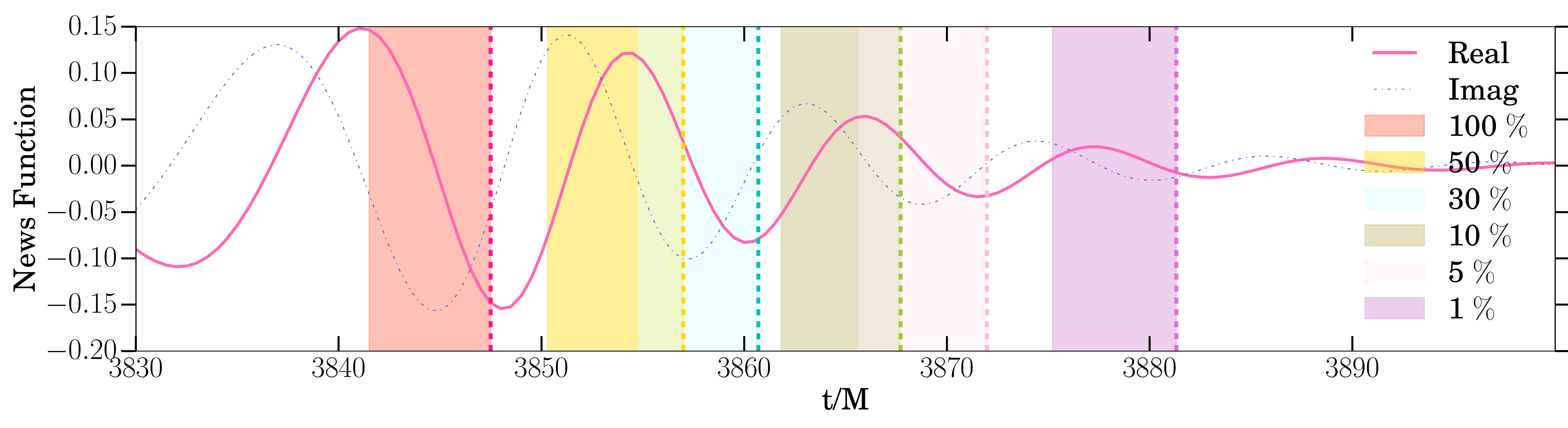} }   
\caption{The concomitant decrease of all of our Kerrness measures. The dashed lines indicate the time at which all the measures decay to at least the indicated $\%$ of peak. The bands color the region in which different measures decrease to the indicated $\%$ of peak. Notice that there is about half a cycle spread in each of these bands. Therefore, the dashed lines provide a conservative idea of the validity of the choice of the start time for data analysis. We have specifically included the spread of these bands as a quantifier of error bounds in the statements of validity made further in this paper. Furthermore, one could shrink the right boundary of these shaded bands if one combines the Kerrness measures with appropriate weights based on their sensitivity to the spacetime curvature and the final remnant's effective potential. }
\label{fig:PercentOnTheWave}
\end{figure*}
}

\newcommand{\ampmap}{%
\begin{figure}[!htbp]
\subfloat{\includegraphics[width=\columnwidth, keepaspectratio]{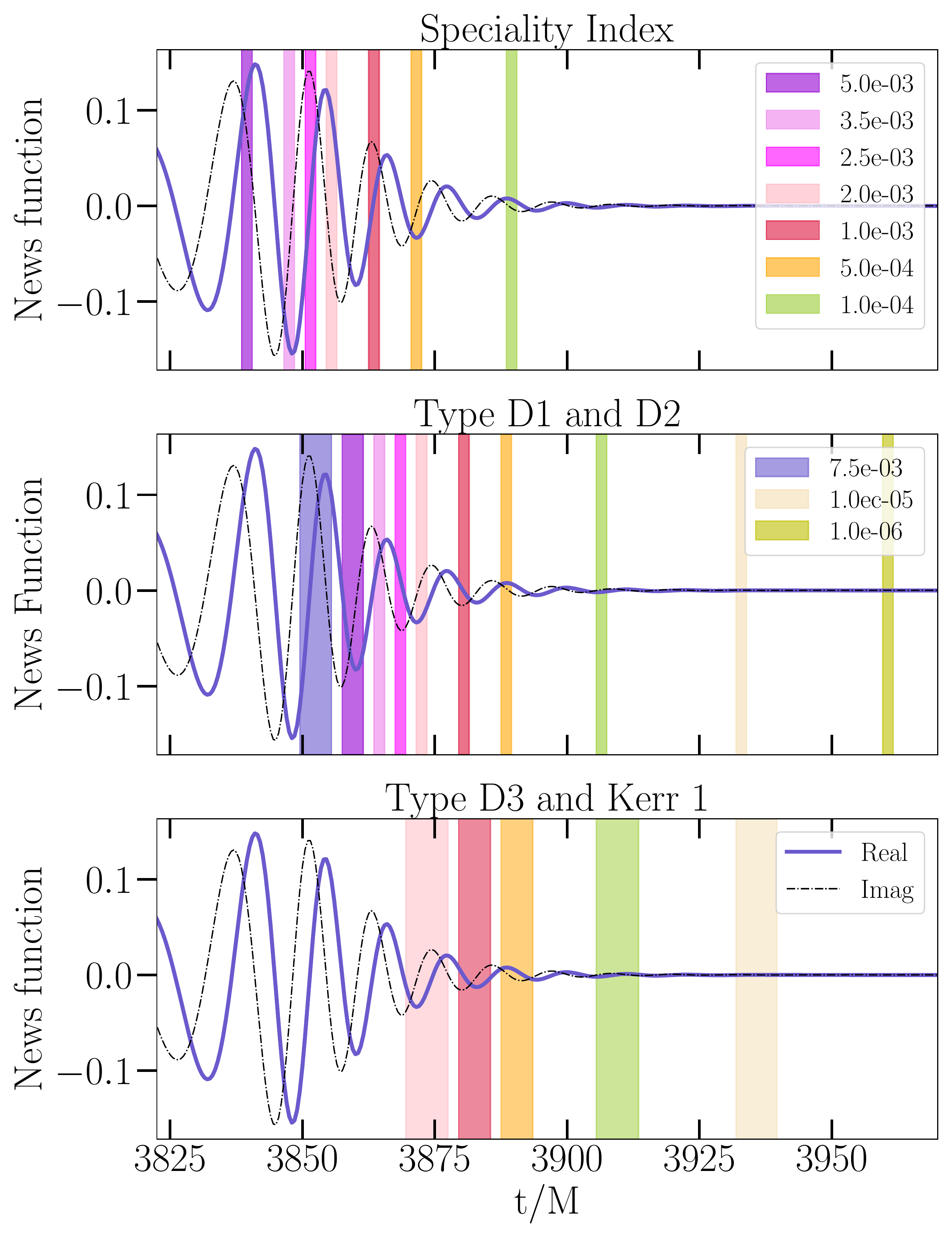}}
\caption{Mapping the inferred perturbation amplitude close to the BH onto the news function. The \textbf{top panel} shows the spread in the crossing times computed using just the Specialty Index, the \textbf{middle panel} uses only the algebraic measures and the \textbf{bottom panel} utilizes only the geometric measures. Notice that amplitudes larger than $2 \times 10^{-3}$ do cross the post-merger timeslices when computed using the geometric measures and that the crossing time spreads in them are relatively large, suggesting a difference in the symmetry of a perturbed Kerr metric and the post-merger BBH spacetime. However, this does not seem to be reflected when we just consider the algebraic measures as they have a relatively small spread in the crossing time. The spread in the crossing time of the Specialty Index is equal to the sampling rate.}
\label{fig:crossingtime}
\end{figure}}

\newcommand{\SNRdp}{%
\begin{figure}[!htbp]
\subfloat{\includegraphics[width=\columnwidth, keepaspectratio]{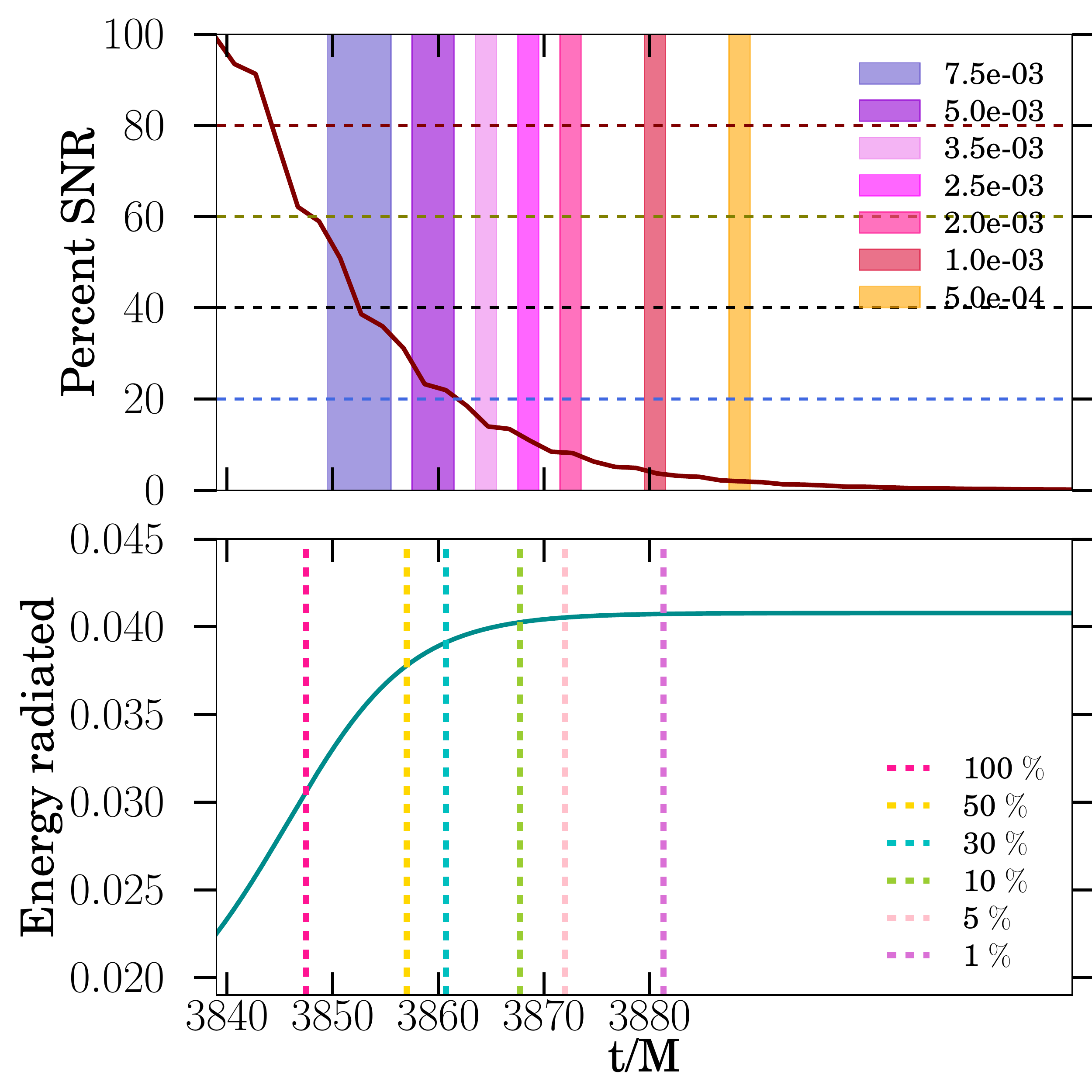} } 
\caption{The\textbf{ top panel} of this figure shows the percentage decrease of SNR from the peak value. The $\%$ SNR is set to $100$ at $t_{\mathrm{merger}}$. For this plot, we evaluate Eq.~\eqref{eq:SNR} with varying lower bounds for the integration. The dashed horizontal lines correspond to $\{80, 60, 40, 20\} \%$ SNR. On the same plot, we mark the perturbation amplitude bands for a direct comparison between perturbation amplitude and statistical error. Notice that by the time the perturbation amplitude near the BH decreases by an order of magnitude, there is only a few percent of SNR left in the signal, emphasizing the sharp trade-off between the systematic biases arising from modeling the post-merger as perturbed Kerr and the statistical uncertainty arising due to exponentially decay of signal amplitude.  The \textbf{bottom panel} shows the total energy radiated in units of $M$ during the merger-ringdown. This is calculated by integrating Eq.~\eqref{eq:EnergyRad}. Again, we have plotted the concomital percentage decrease of the Kerrness measures from their peak values for an easy comparison between the statistical and systematic errors associated with the choice of the start time of ringdown. In particular, the constant settling in the total radiated energy occurs between the time when the Kerrness measures have decayed to $5-1 \%$ of their peak values, implying that at these times the GW is very weak in amplitude.  }
\label{fig:PercentageSNR}
\end{figure}
}

\newcommand{\postIt}{%
\begin{figure*}
\subfloat{\includegraphics[width=0.5\textwidth]{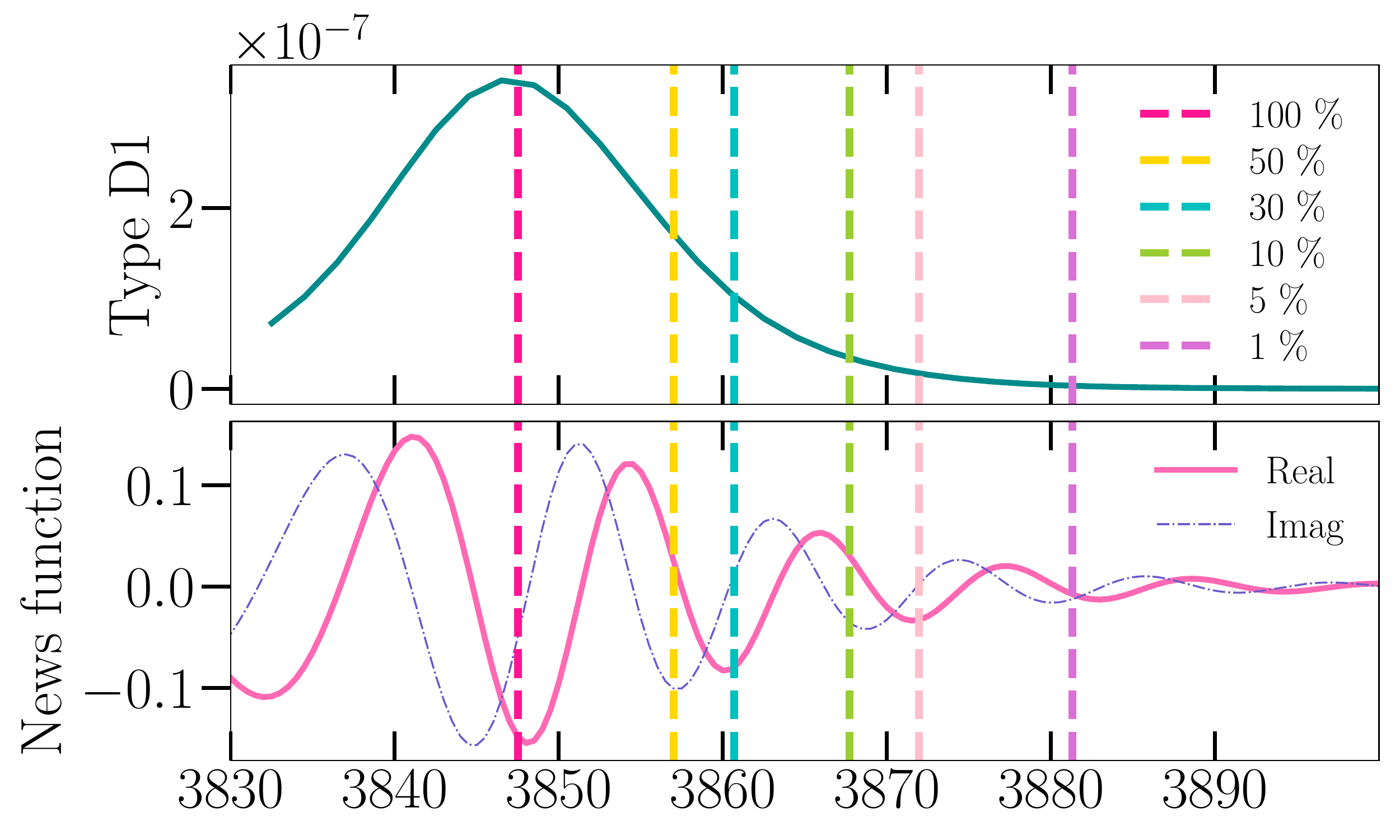} }
\subfloat{\includegraphics[width=0.5\textwidth]{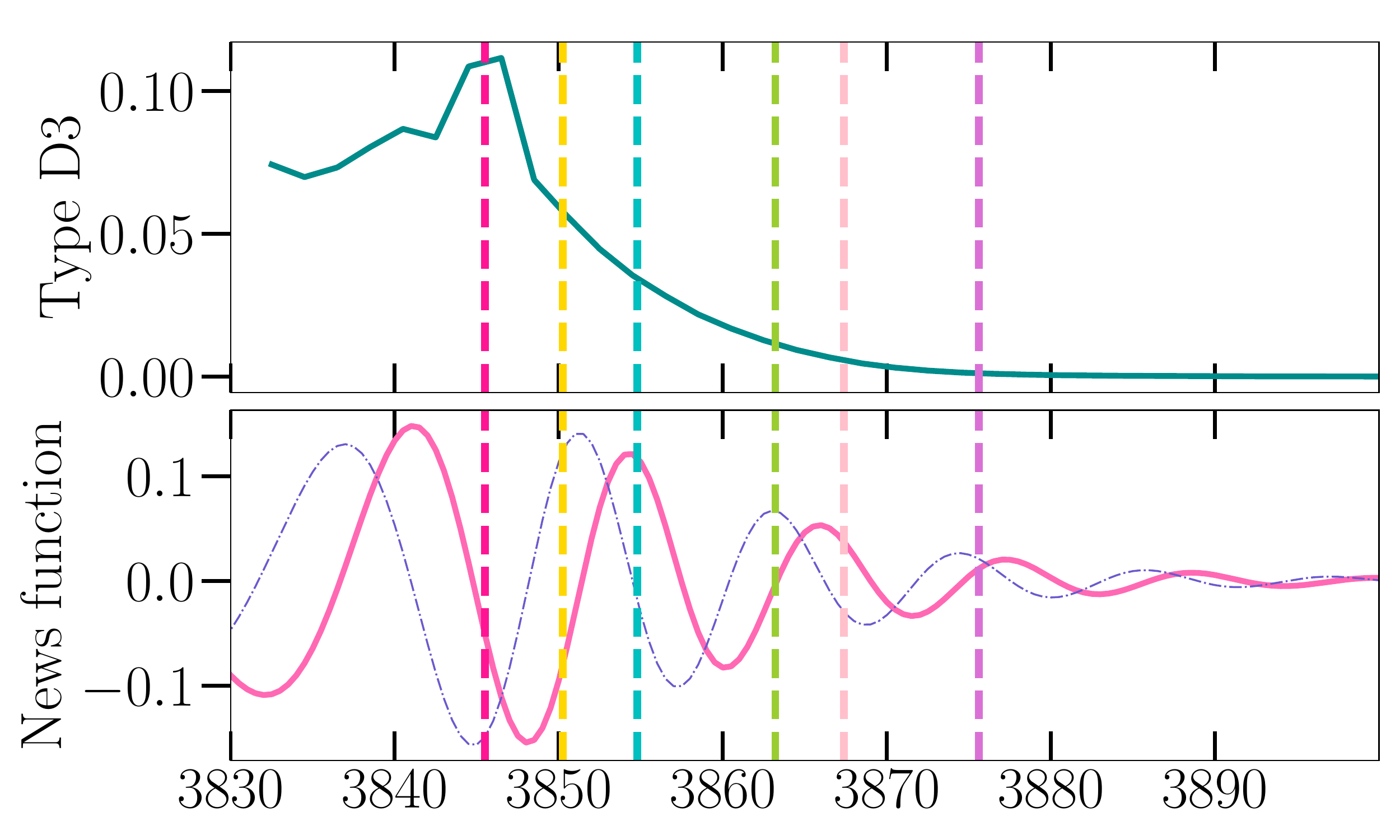} }  \\[-0.3cm]
\subfloat{\includegraphics[width=0.5\textwidth]{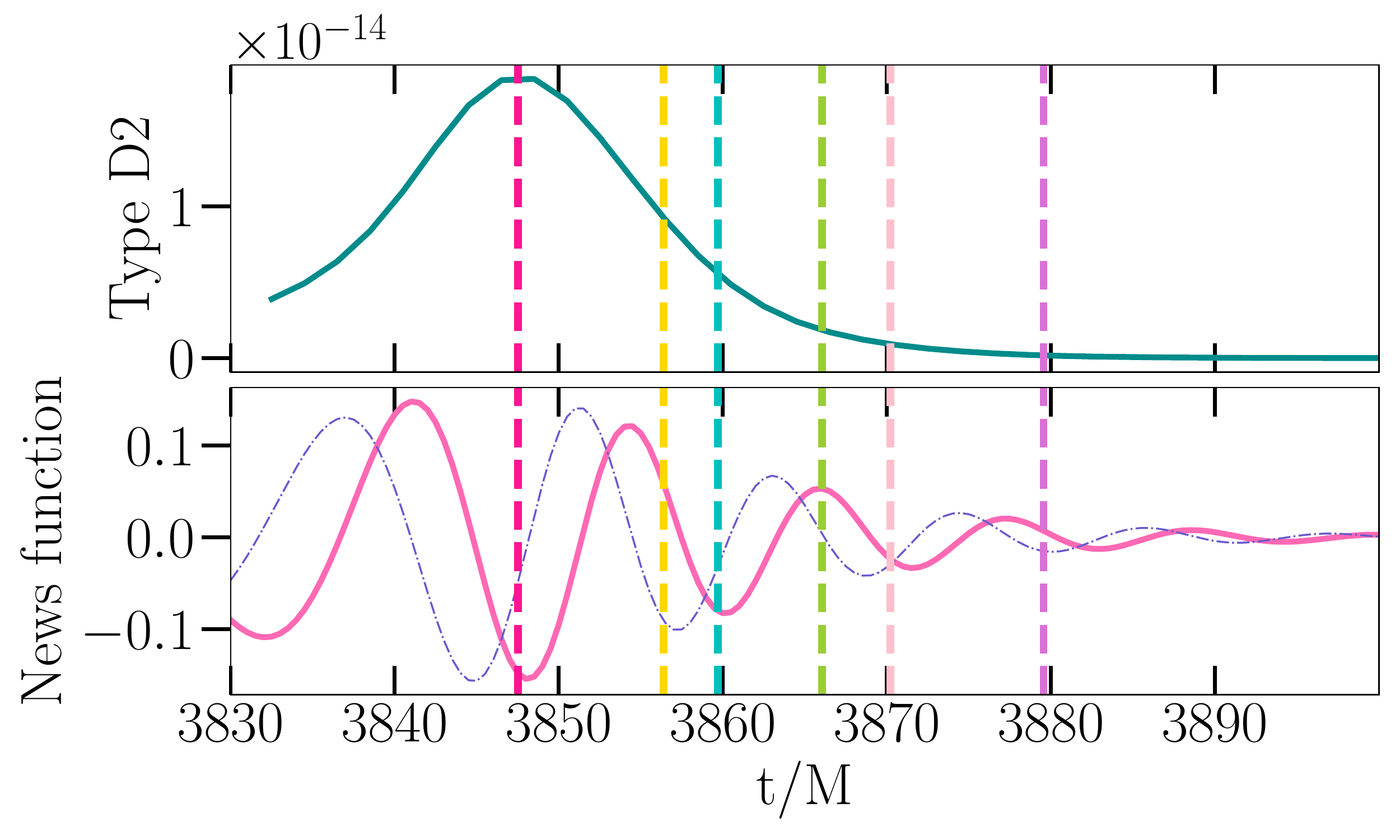} }
\subfloat{\includegraphics[width=0.5\textwidth]{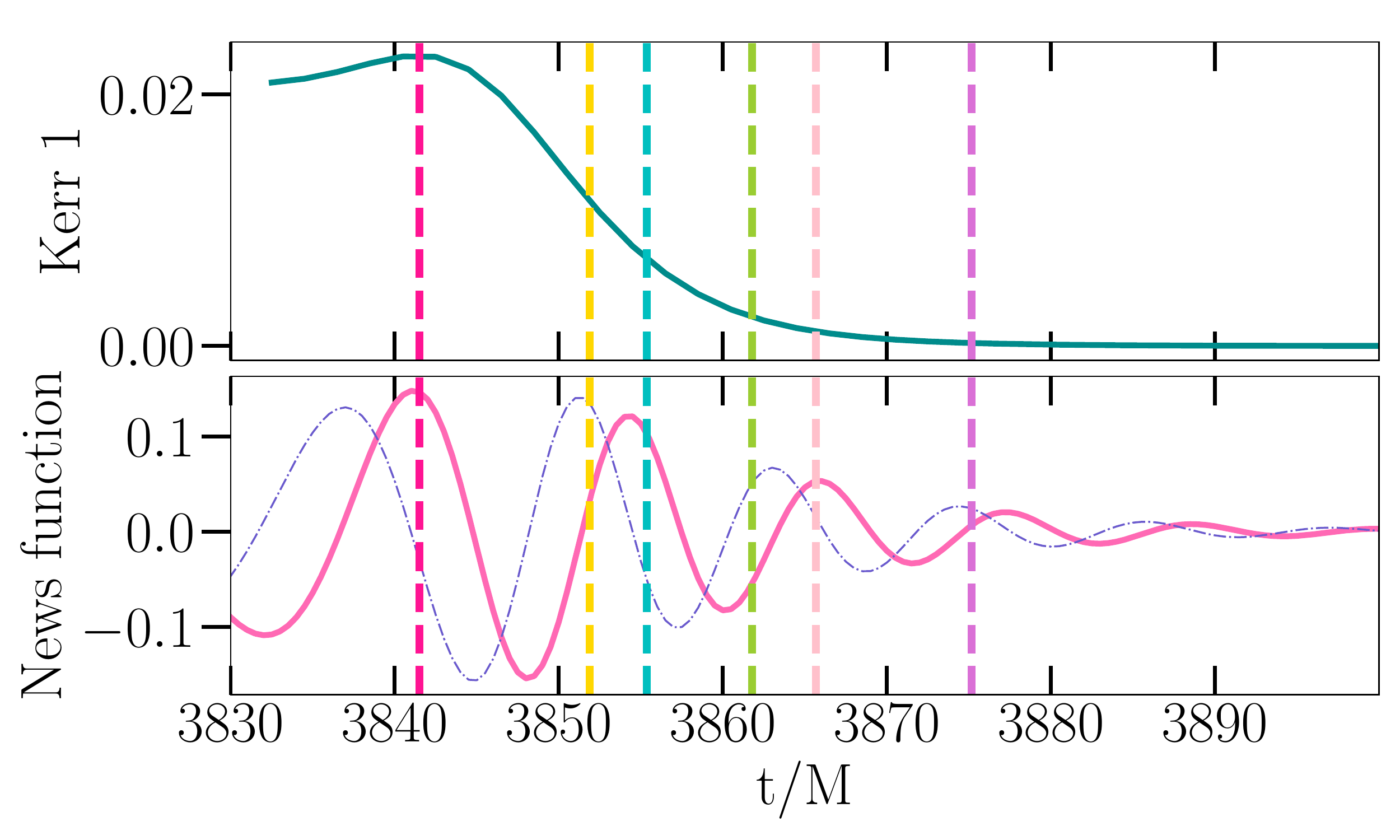} }
\caption{Connecting the Kerrness measures in the strong-field to dynamics at $\fni$ using the procedure described in Sec.~\ref{sec:AFQ_implementation} on the BBH post-merger. The \textbf{left panels} map the algebraic measures and the \textbf{right panels} map the geometric measures on to the news function. The \textbf{top panel} within each subplot corresponds to a Kerrness measure in the strong-field, while the \textbf{bottom panel} shows the news function at $\fni$. The dashed lines of different colors indicate the $\%$ decrease from the peak value of the respective Kerrness measures. The horizontal axis corresponds to the simulation coordinate time induced on the news function extracted from a world tube radius of $R=128\,M$. Furthermore, unlike the strong-field result plots that aim at rigorous characterization of isometry to Kerr, here we aim at providing insight into validating the start time of ringdown for data analysis. Therefore, these plots are on linear scale as opposed to logarithmic scale. Notice that the curves on the left panel decay more slowly than those on the right; Type D 1 is the slowest to decay, closely followed by Type D 2. Also, recall that we cannot compare the magnitude of the top part of each of these panels as they are dimensionally different. }
\label{fig:PostItPanel}
\end{figure*}}

\newcommand{\phiplot}{%
\begin{figure}[!htbp] \subfloat{\includegraphics[width=\columnwidth]{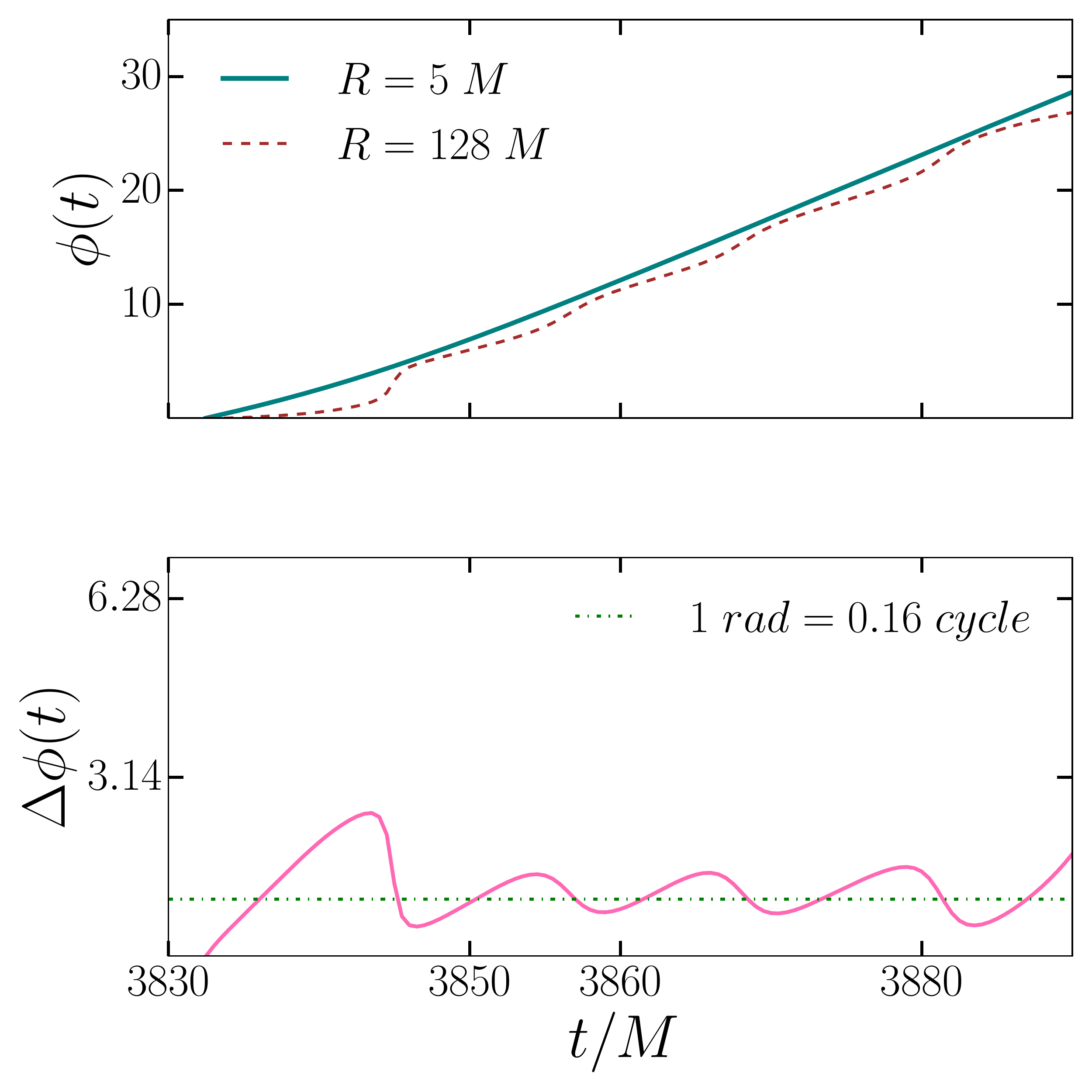} } 
\caption{The phase discrepancy between the news function extracted from a worldtube radius of $R=5\,M$ and $R= 128\,M$. The news functions are aligned to maximize the overlap. The \textbf{top panel} presents the phase evolution of the news function for each extraction radius. The \textbf{bottom panel} shows the fractional difference defined as $\phi_{128} - \phi_{5}$. Notice that the phase difference is significant at the very beginning but quickly decreases to an acceptable level for our analysis. We notice that the phase difference oscillates about 1 radian, indicating the level of error we introduce by - a) not performing the final gauge transformation, b) imposing no-ingoing condition for CCE.  }
\label{fig:phasediffcomb}
\end{figure}}

\newcommand{\cartoon}{%
\begin{figure}[!htbp]
\centering
\subfloat{\includegraphics[width=\columnwidth]{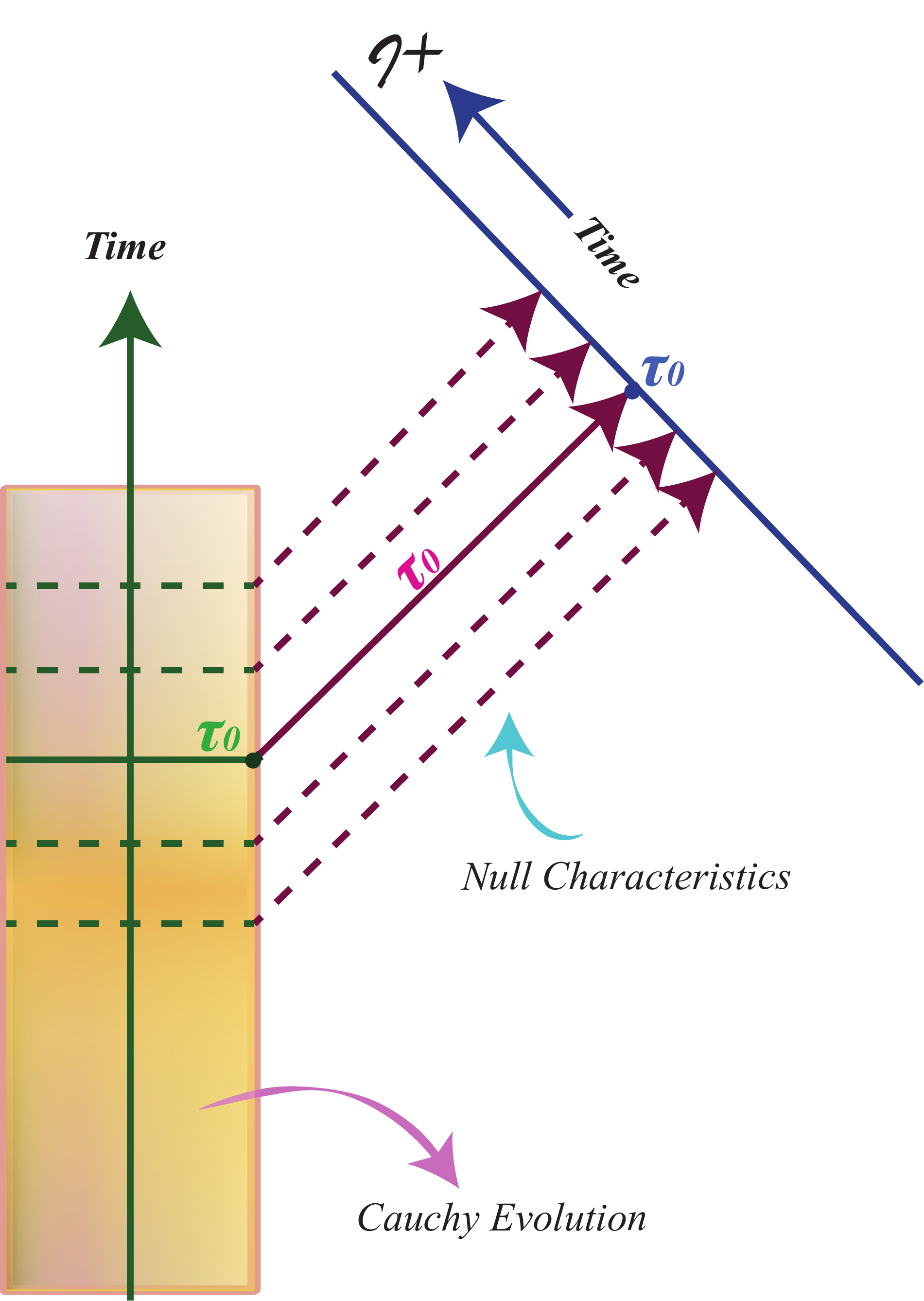} }
\caption{Prescription for connecting the strong-field information to the asymptotic frame dynamics. The colored cylinder represents the region of spacetime that is evolved by the Cauchy code. The vertical green line within the cylinder indicates the direction of coordinate time. The horizontal lines represent time slices. The details of the location of time slices depend on the gauge choice. The pink boundary of the cylinder depicts the worldtube from where the CCE is performed. The purple lines with unit slope illustrate the null characteristics along which the information on the worldtube is propagated to (the solid blue line) $\fni$. In our procedure of associating information in the source frame with the asymptotic frame, we identify all the points along a characteristic by an equivalence. The solid green line in the cylinder acts as a source to the waveform feature at $\tau_0$ observed at $\fni$.}
\label{fig:Cartoon}
\end{figure}
}

\newcommand{\tgr}{%
\begin{figure}[!htbp]\subfloat{\includegraphics[width=\columnwidth]{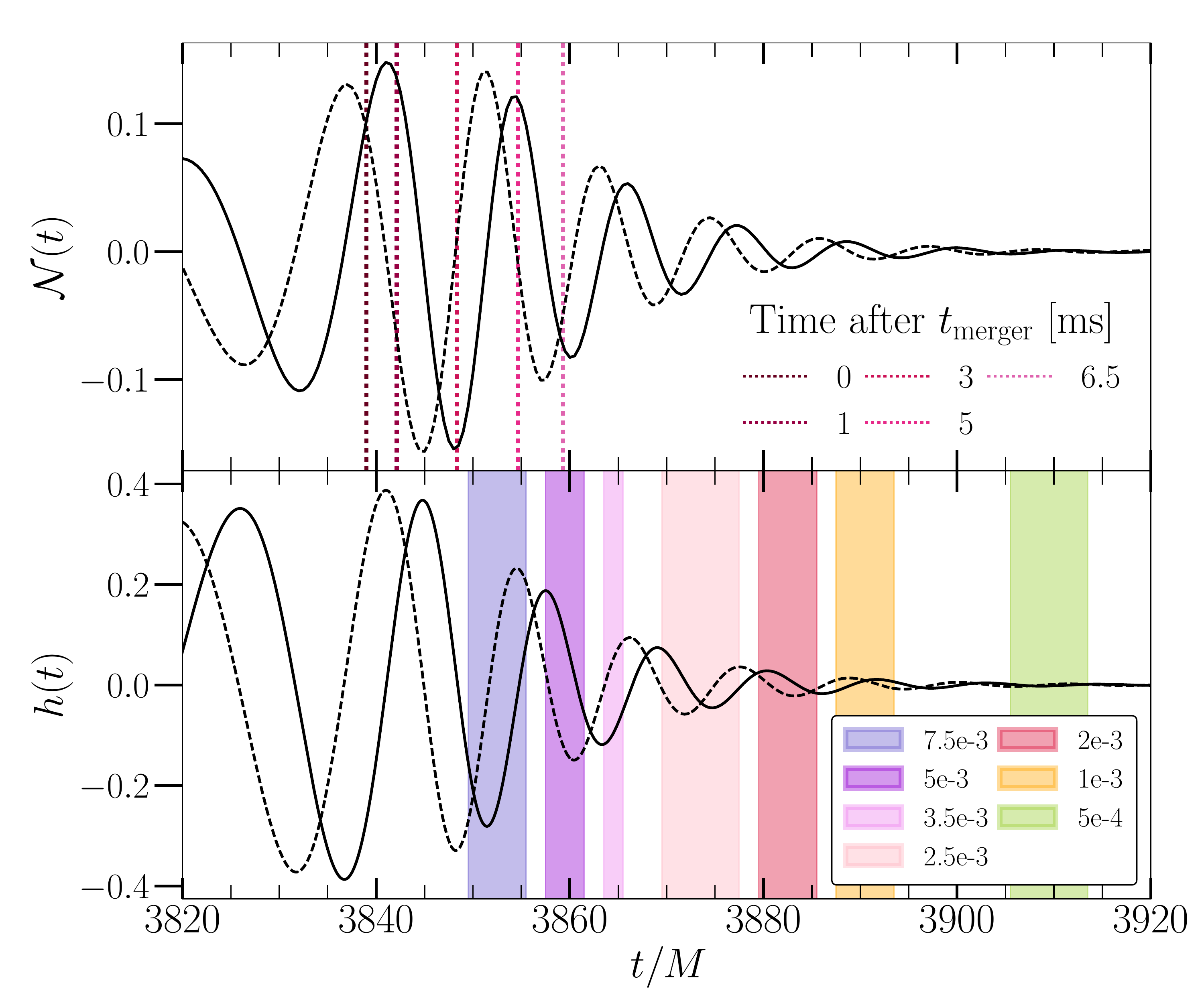} } 
\caption{Comparison of the times chosen in the testing GR study of GW150914~\cite{TheLIGOScientific:2016src}. Here, we make statements about their validity to perform tests that rely on the perturbative nature of the BH. Specifically, we propose that a plot of this nature be done for future detections, especially if the SNR is high, to gain an insight into the inferred strong-field perturbation amplitudes corresponding to different choices of ringdown start time. The dotted line in the \textbf{top panel} shows different choices of start time for performing tests on the detector data. The \textbf{bottom panel} shows what each time choice corresponds to in the simulation gauge. Although a practical choice of start time to perform tests like no-hair theorem tests should be decided based on the interplay between the statistical and systematic uncertainty, a plot of this nature gives significant understanding of the results of such tests. For instance, in the case of GW150914, had the signal been much louder than what we observed, this plot suggests that we \textit{could} get biased results due to large inferred perturbation amplitude in the strong-field leading to errors in modeling the post-merger as a perturbed BH at 3 ms.}
\label{fig:TGR}
\end{figure}
}

\newcommand{\ConvergenceTestFigure}{%
\begin{figure}[!htbp]
\subfloat{\includegraphics[width=0.45\textwidth]{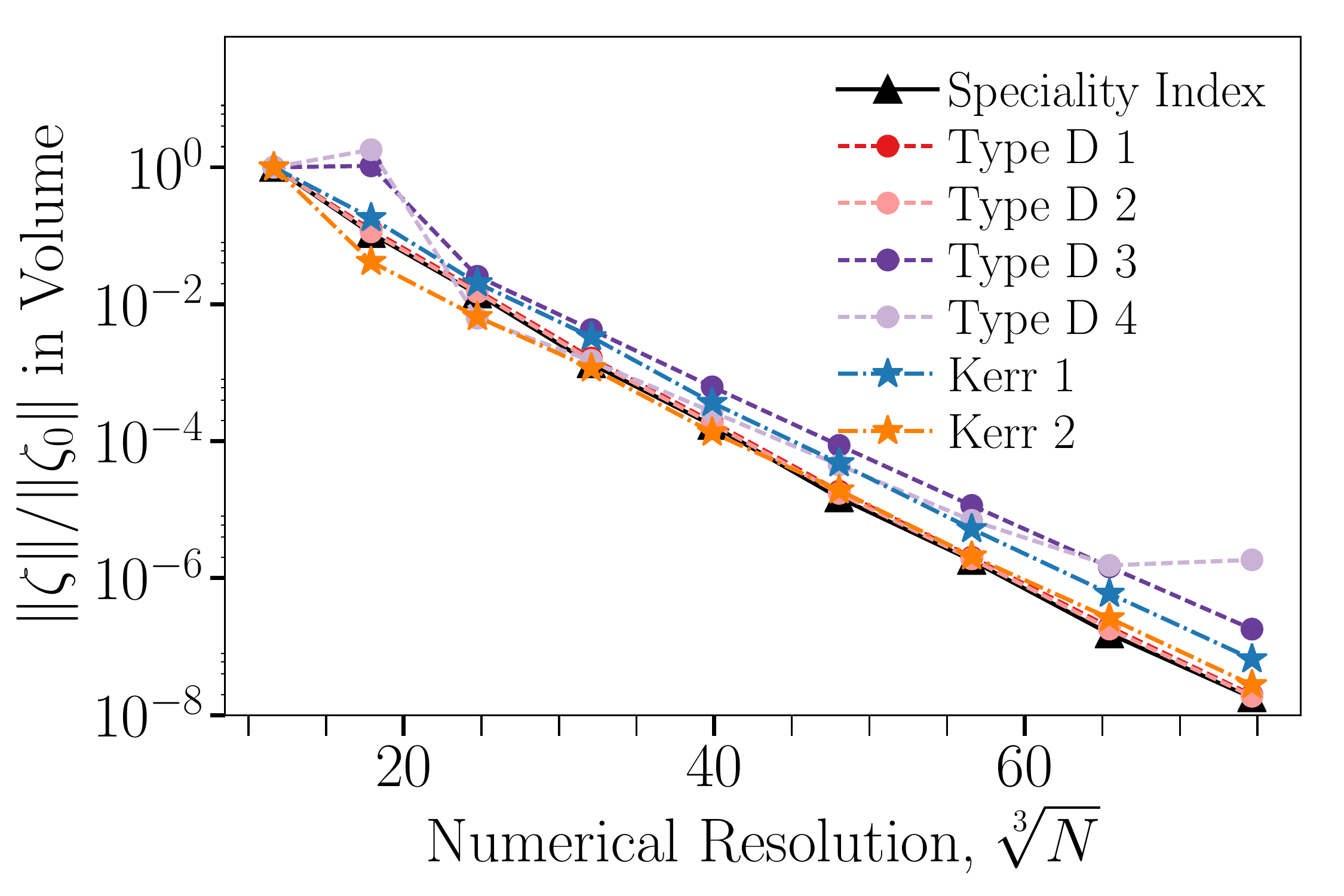} } 
\caption{Convergence of Kerrness measures on a numerical BH in Kerr-Schild coordinates with 
dimensionless spin $\chi = (0.2, 0.3, 0.4)$. We observe exponential convergence towards the theoretical value of zero with numerical resolution. For each measure $\zeta$, we present $\|\zeta\|/\|\zeta_0\|$, the L2 norm over the spatial slice normalized by the L2 norm of the lowest resolution. The resolution is expressed $\sqrt[3]{N}$, where $N$ is the number of spectral collocation points in the domain.
}
\label{fig:ConvergenceTest}
\end{figure}
}

\newcommand{\EnvelopesFigure}{%
\begin{figure}[!htbp]
\subfloat{\includegraphics[width=\columnwidth]{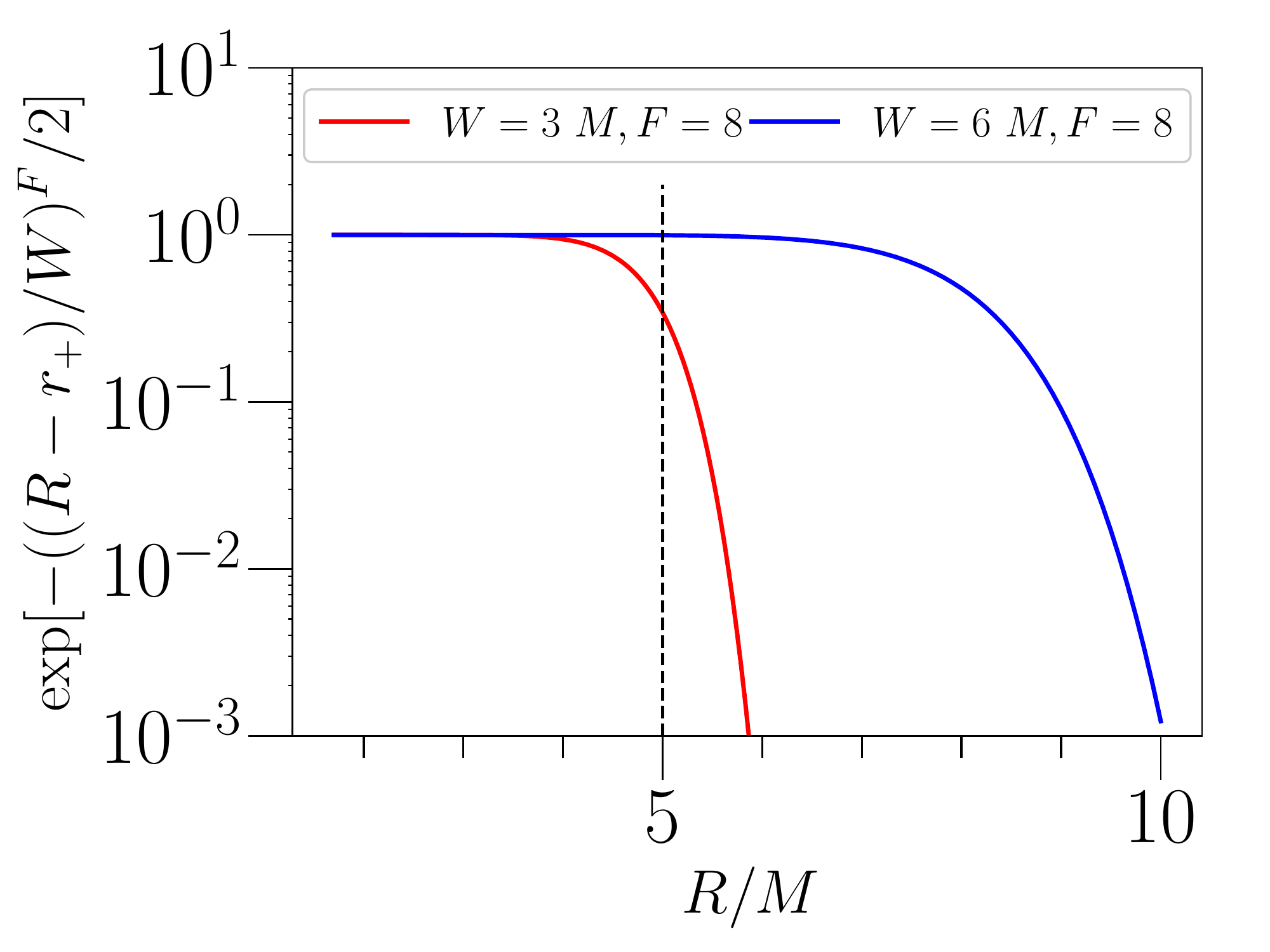} } 
\caption{Envelope function from Eq.~\eqref{eq:envelope}, for various width and falloff parameters, $\{W, F\}$. We show how the envelope parameters affect an  extraction radius of $R=5\,M$ (marked by the dashed black line). For our chosen values of $\{W = 6\,M, F = 8\}$, the envelope is at $\sim 1$ and $R=5\,M$, while for $\{W = 3\,M, F = 8\}$, the envelope affects the perturbation amplitude at $R=5\,M$. We have checked that using a smaller envelope does not change the qualitative behavior of our results.}
\label{fig:Envelopes}
\end{figure}
}

\newcommand{\KerrPertAmplitudeFigure}{%
\begin{figure}[!htbp]
\subfloat{\includegraphics[width=\columnwidth]{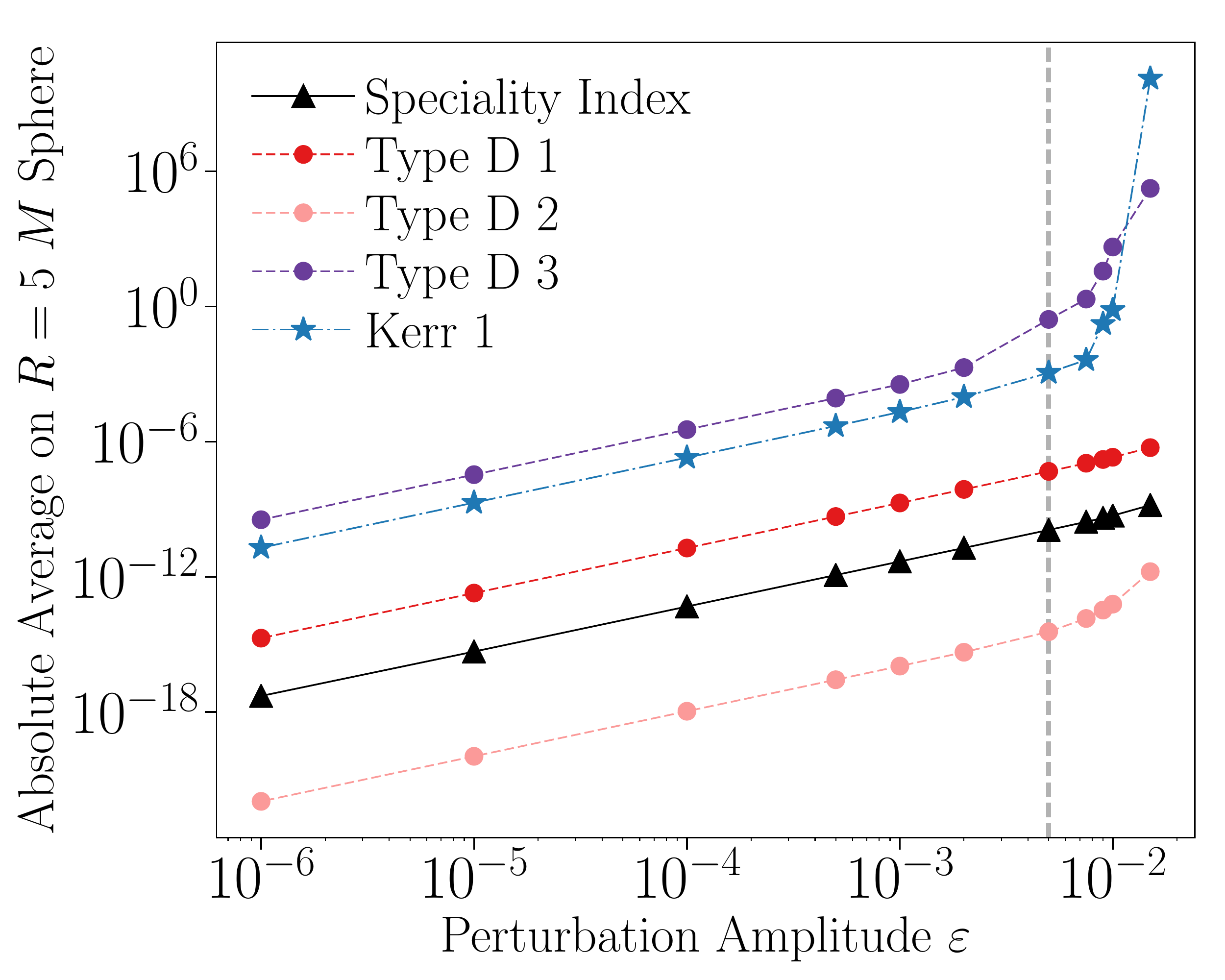} } 
\caption{Behavior of absolute Kerrness measures with perturbation amplitude $\varepsilon$. We compute this on an $l=m=2$ QNM perturbed Kerr BH with the same mass and spin as the final remnant in the BBH simulation we consider in this paper. We average each measure on a coordinate 2-sphere of $R=5\,M$. The behavior is initially quadratic with $\varepsilon$ for all measures. At larger amplitudes $\varepsilon \geq 5 \times 10^{-3}$, Type D 2, D 3, D 4, and Kerr 1 show higher-power dependence, and hence non-linearity. We show this  $\varepsilon_\mathrm{crit} \sim 5 \times 10^{-3}$ by a dashed vertical line. The lines between the points are only used to visually connect them (rather than being fits).
}
\label{fig:KerrPertAmplitude}
\end{figure}
}

\newcommand{\HorizonDataFigure}{%
\begin{figure}[!htbp]
\subfloat{\includegraphics[width=\columnwidth]{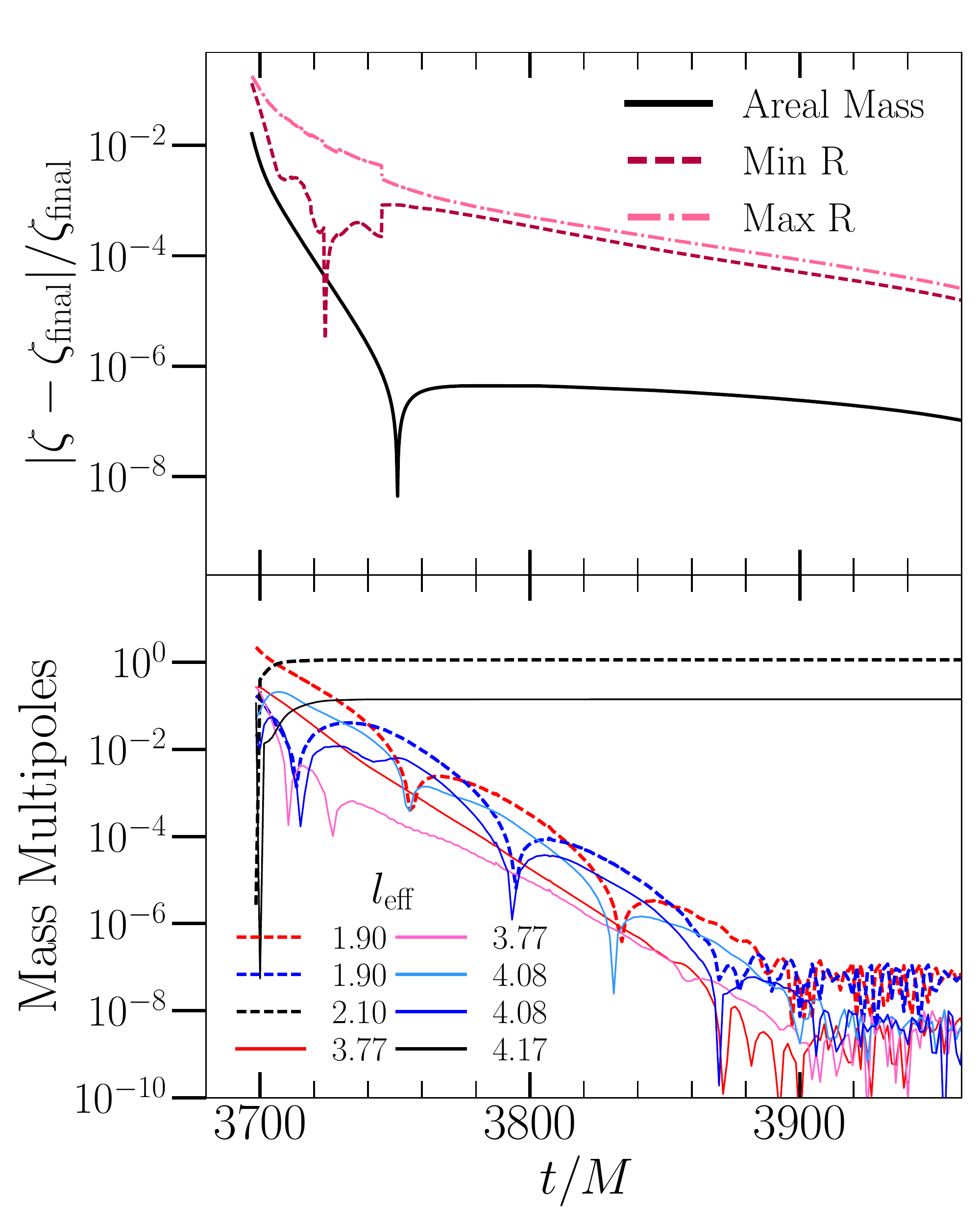} } 
\caption{Settling of the post-merger AH as a function of coordinate time. The \textbf{top panel} shows the areal mass quickly attaining a constant value and the minimum and maximum radii $R$ of the horizon exponentially settling to final values. Each quantity $\zeta$ is presented as $|\zeta  - \zeta_\mathrm{final} |/\zeta_\mathrm{final}$ where $\zeta_\mathrm{final}$ is the value at the final time of the simulation. The \textbf{bottom panel} shows the behavior of the initially excited AH mass multipoles, labeled by the $l_\mathrm{eff}$ given in Eq.~\eqref{eq:leff} at the final time. The initially excited quadruple moments ($l_\mathrm{eff} \sim 2$) are shown by the dashed lines, while the initially excited hexadecupole moments ($l_\mathrm{eff} \sim 4$) are shown by the solid lines. As discussed in the text, two of the quadropule moments and four of the hexadecupole moments, as well as the $l \sim 1$ and $l \sim 3$ moments immediately vanish due to symmetry. Thus, we do not plot them in this figure. The excited multipoles either exponentially decay or reach constant values consistent with the values expected for Kerr~\cite{Owen:2009sb}.
}
\label{fig:HorizonData}
\end{figure}
}

\newcommand{\RainbowFigure}{%
\begin{figure*}
\subfloat{\includegraphics[width=\textwidth]{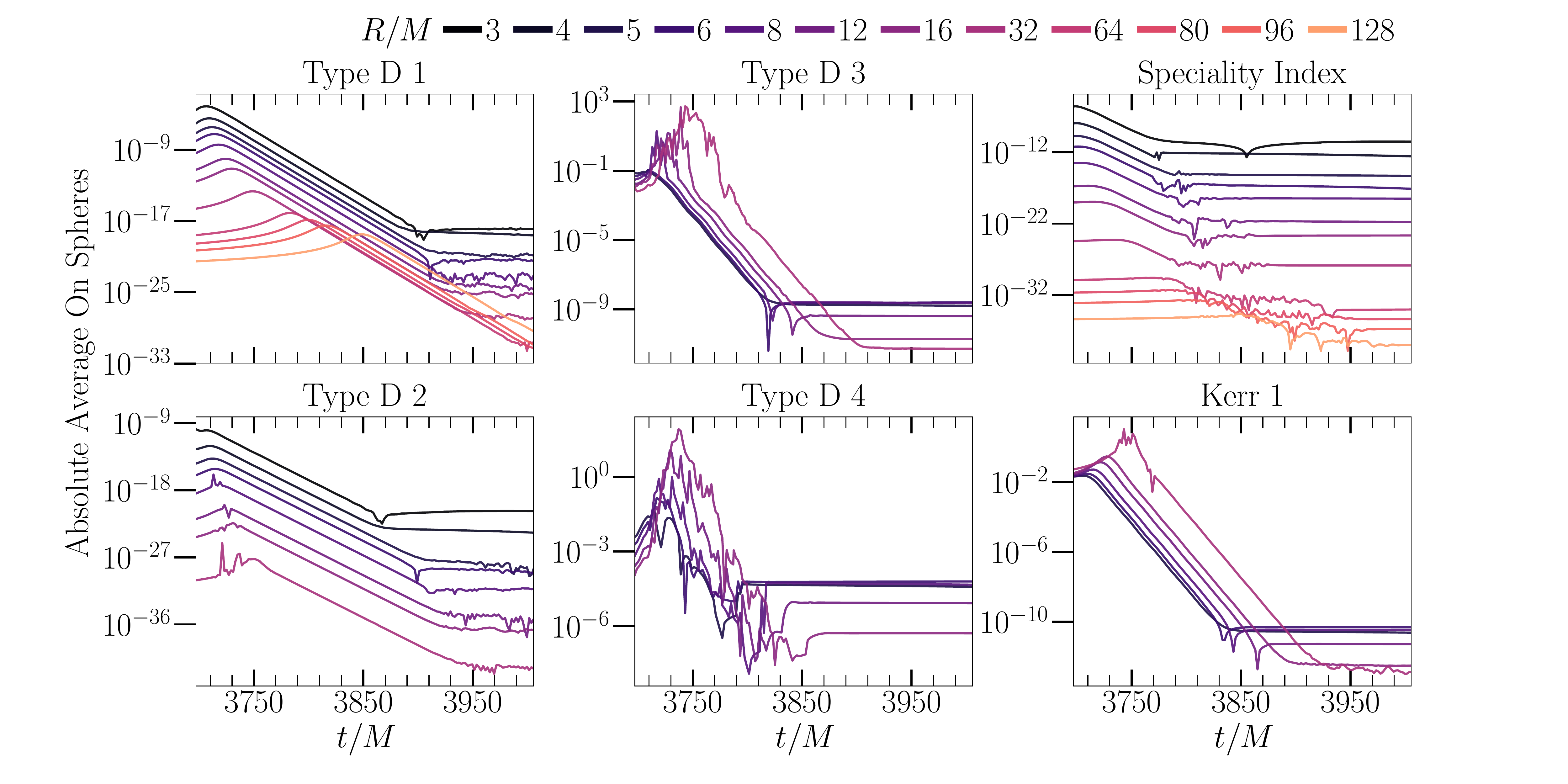} } 
\caption{Behavior of absolute Kerrness measures with coordinate time on BBH post-merger spacetime. The measures are averaged on a variety of concentric nested coordinate 2-spheres of radii $R$ around the BH, as indicated by the colors. Larger values \textit{within each subplot} mean that the 2-sphere is farther from being locally isometric to Kerr. For measures that involve higher-order numerical derivatives, we present the results only at radii  where they are at least somewhat well resolved. All plots, however, include $R = 5\,M$, the radius we use to map Kerrness onto the waveform. Type D 4 is particularly noisy, as it contains the highest number of numerical derivatives. The measures exponentially decay as the spacetime approaches Kerr, ultimately reaching a numerical noise floor. We observe that the peak of each measure moves outwards with radius, indicating propagation of non-Kerrness.}
\label{fig:Rainbow}
\end{figure*}
}

\newcommand{\NoiseFloorFigure}{%
\begin{figure}[!htbp]
\includegraphics[width=\columnwidth]{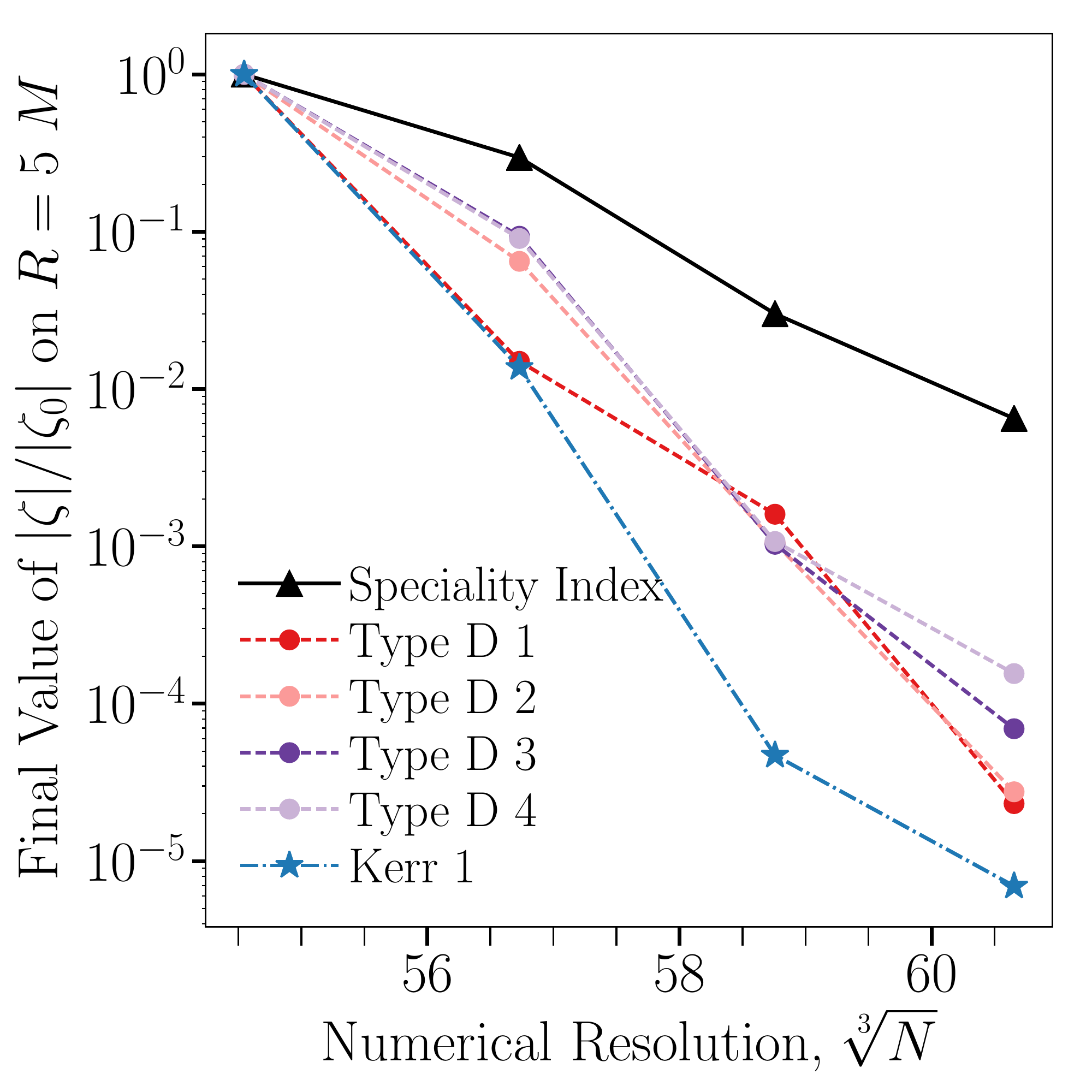}
\caption{Exponential convergence of the noise floor of each Kerrness measure on the final timestep of the BBH simulation. Each measure $\zeta$ is presented as an average over a 2-sphere of $R=5\,M$ (where the measures have settled to a noise floor), normalized by $|\zeta_0|$, the average of the lowest resolution. The resolution is indicated by $\sqrt[3]{N}$, where $N$ is the number of spectral collocation points. The convergence to zero shows that the  noise floor observed in Fig.~\ref{fig:Rainbow} is a numerical noise floor, rather than real a physical artifact. We have also testing this convergence behavior on a 2-sphere $R=5\,M$ and verified that the behavior is consistent (although more noisy).}
\label{fig:NoiseFloor}
\end{figure}
}

\newcommand{\KerrTwoFigure}{%
\begin{figure}[!htbp]
\subfloat{\includegraphics[width=\columnwidth]{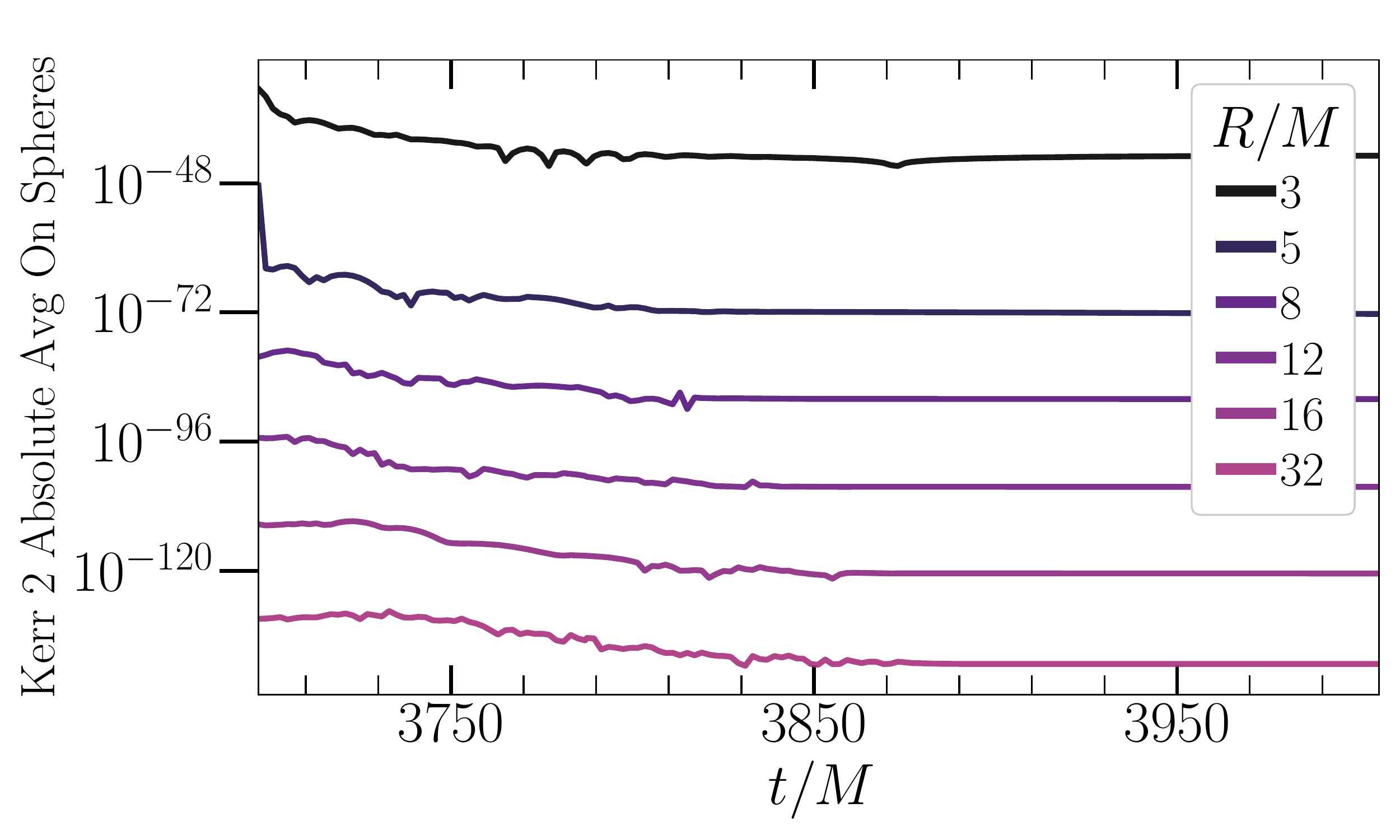} } 
\caption{Kerr 2 measure throughout the post-merger BBH simulation, averaged on a variety of coordinate 2-spheres of radius $R$. The values remain relatively constant and low, indicating that no NUT charge is gained during ringdown.}
\label{fig:KerrTwo}
\end{figure}
}

\newcommand{\SwirlFigure}{%
\begin{figure*}
\subfloat{\includegraphics[width=0.8\textwidth]{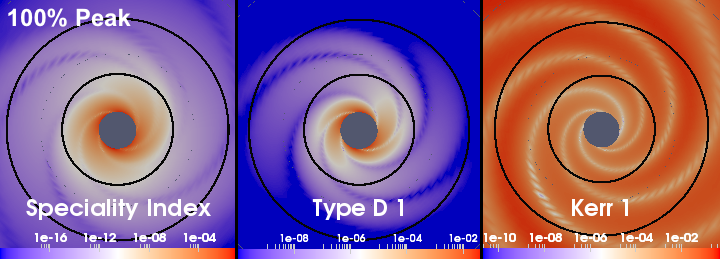} } \\[-0.05cm]
\subfloat{\includegraphics[width=0.8\textwidth]{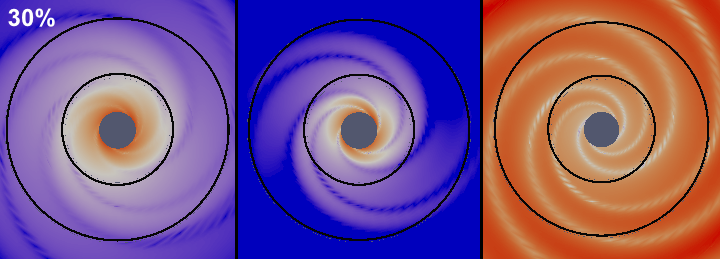} } \\[-0.05cm]
\subfloat{\includegraphics[width=0.8\textwidth]{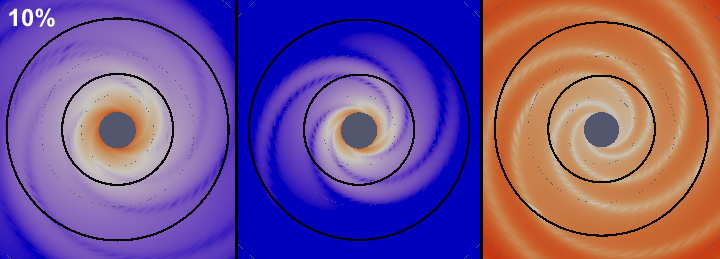} } \\[-0.05cm]
\subfloat{\includegraphics[width=0.8\textwidth]{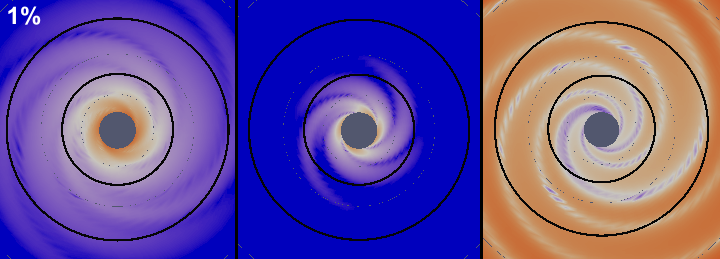} } \\[-0.05cm]
\caption{Absolute Kerrness measures on slices of the BBH post-merger spacetime. The data is presented in the equatorial plane, with the gray region corresponding to the excised BH. The black circles correspond to coordinate radii $R=5\,M$ and $R=10\,M$. The columns correspond to Speciality Index, Type D 1, and Kerr 1, and the rows (from top to bottom) correspond to coordinate times at which the each measure at $R=5\,M$ achieves $100\%$, $30\%$, $10\%$, and $1\%$ of the combined peak value.  The quadrupolar pattern (with $|m| = 2$) in all three measures is consistent with the dominant quadrupolar radiation (recall that these are absolute measures, and hence do not distinguish between positive and negative values). Notice that the algebraic measures---Speciality Index and Type D 1---settle outward-in, whereas Kerr 1, a geometric measure, settles inward-out. Additionally, the structures in the measures are visible even at $1\%$ of the peak value. We can compare these measures to $\Psi_4$ (in Figs.~\ref{fig:psi41B}~and~\ref{fig:psi42B}) to infer their sensitivity to the spacetime curvature features.}
\label{fig:Swirl}
\end{figure*}
}

\newcommand{\PsiFourComparisonFigure}{%
\begin{figure}[!htbp] 
\subfloat[$\Psi_4$ on $\varepsilon = 7.5 \times 10^{-3}$ single BH]{\includegraphics[width=0.5\columnwidth]{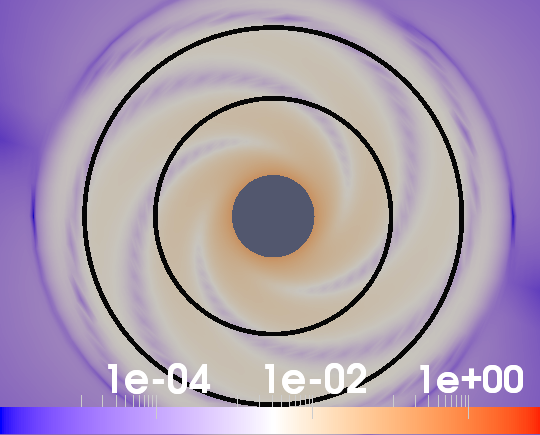} \label{fig:psi41A}}
\subfloat[$\Psi_4$  for BBH at $\varepsilon = 7.5 \times 10^{-3}$ crossing time]{\includegraphics[width=0.5\columnwidth]{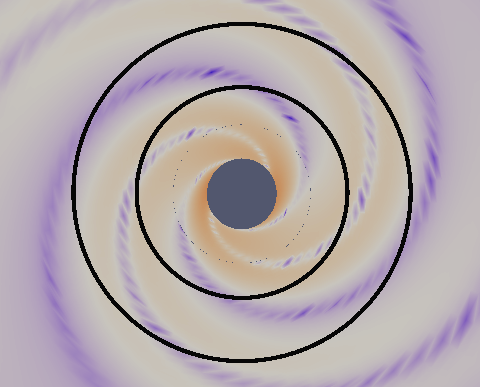} \label{fig:psi41B}}\\ [-0.3 cm]
\subfloat[$\Psi_4$ on $\varepsilon = 10^{-3}$ single BH]{\includegraphics[width=0.5\columnwidth]{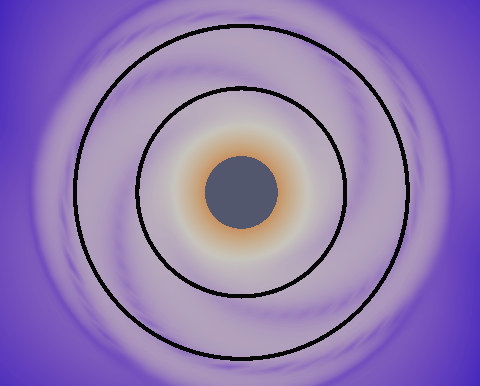} \label{fig:psi42A}}
\subfloat[$\Psi_4$ for BBH at $\varepsilon = 10^{-3}$ crossing time]{\includegraphics[width=0.5\columnwidth]{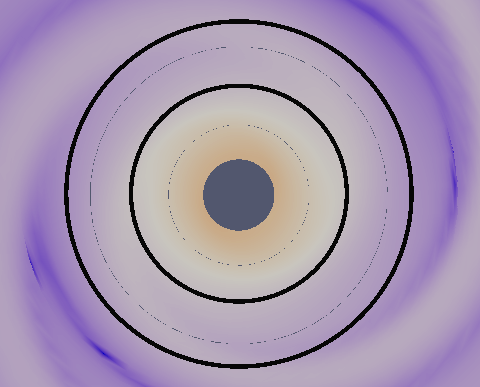} \label{fig:psi42B} }
\caption{$\Psi_4$ in the equatorial plane for both a single BH with an $l=m=2$ perturbation of amplitude $\varepsilon = 7.5 \times 10^{-3}$ and $\varepsilon = 10^{-3}$ (\textbf{left panel}), and for the BBH ringdown (\textbf{right panel}) at times that achieve the same Kerrness as (\textbf{left panel}). For all cases, Kerrness is matched on a coordinate 2-sphere of $R=5\,M$. The two black circles correspond to coordinate radii $R=5\,M$ and $R=8\,M$. The Gaussian envelope of width $R=8\,M$, as described in Fig.~\ref{fig:Envelopes}, can be seen in the plots for the single BH cases. Notice that although the two systems look similar, allowing us to infer the BBH simulation perturbation amplitude, the mapping does have some imperfections.
}
\label{fig:Psi4Comparison}
\end{figure}
}

\newcommand{\CrossingTimeFigure}{%
\begin{figure*}
\subfloat{\includegraphics[width=\textwidth]{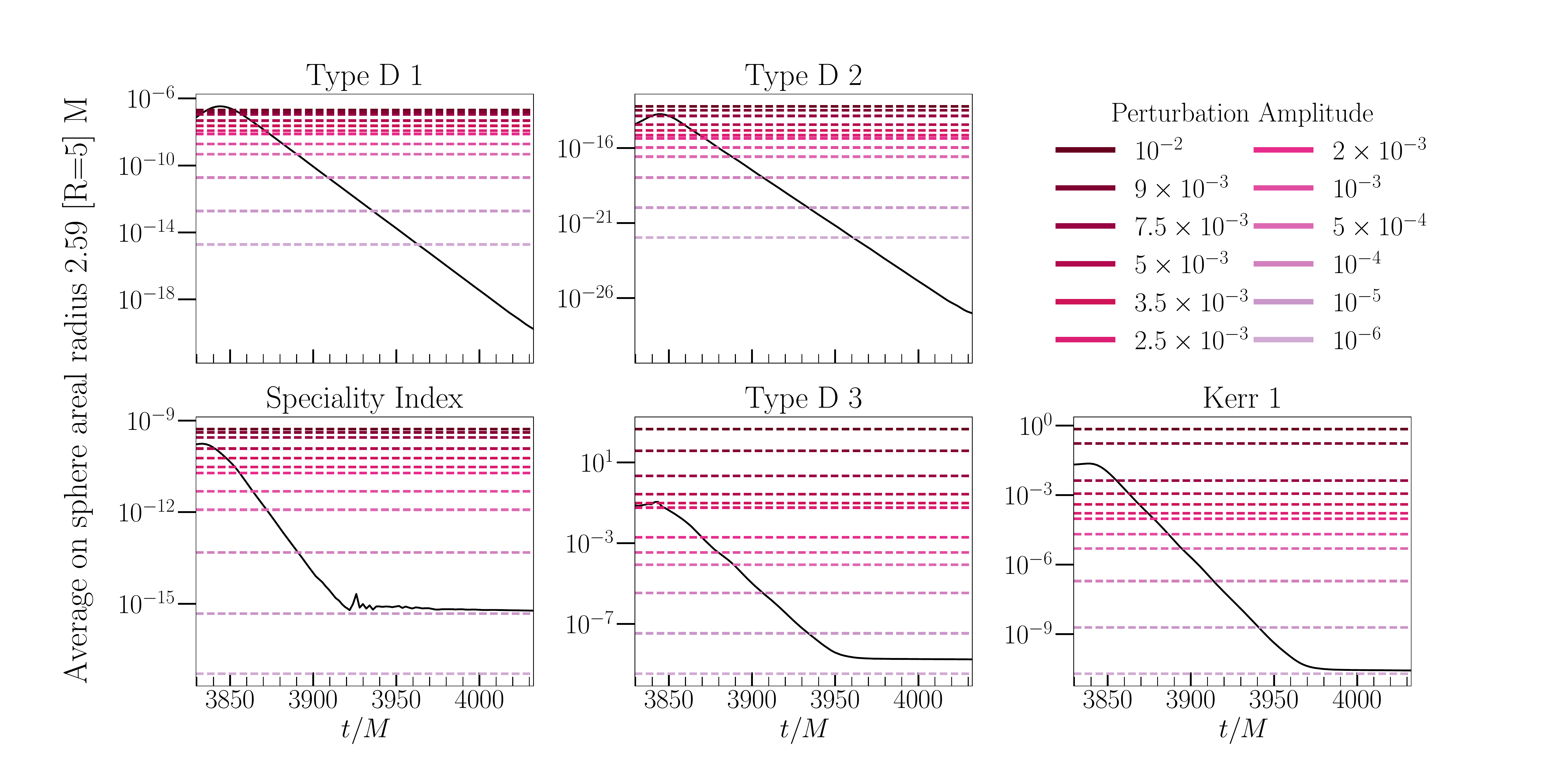} } 
\caption{
Comparison of the Kerrness measures during the BBH post-merger to the values of the Kerrness measures on an $l=m=2$ QNM perturbed Kerr BH of various perturbation amplitudes $\varepsilon$, with the same mass and spin parameters. The measures are averaged on a 2-sphere of coordinate radius $R=5\,M$, which corresponds to comparable areal radii of $\sim 2.59\,M$ in both systems. The measures evaluated on the BBH slices are shown by solid black lines, decaying as a function of time. The Kerrness measures for the perturbed metric are presented as horizontal dashed red lines, one for each $\varepsilon$. The times at which the BBH curves intersect the Kerrness values for a given $\varepsilon$ Kerr perturbation give a scale for the BBH Kerrness measures as the post-merger progresses. These times, known as \textit{crossing times} are then mapped onto the waveform, and used to validate the start time of ringdown. Note that the measures have different crossing times (cf. Fig.~\ref{fig:Psi4Comparison}). The time axes are shifted to agree with the timestamps of the GW at $R= 128\,M$, as explained in Table~\ref{tab:Tshift}.
}
\label{fig:CrossingTime}
\end{figure*}
}

\newcommand{\StrainsFigure}{%
\begin{figure}[!htbp]
\subfloat{\includegraphics[width=\columnwidth]{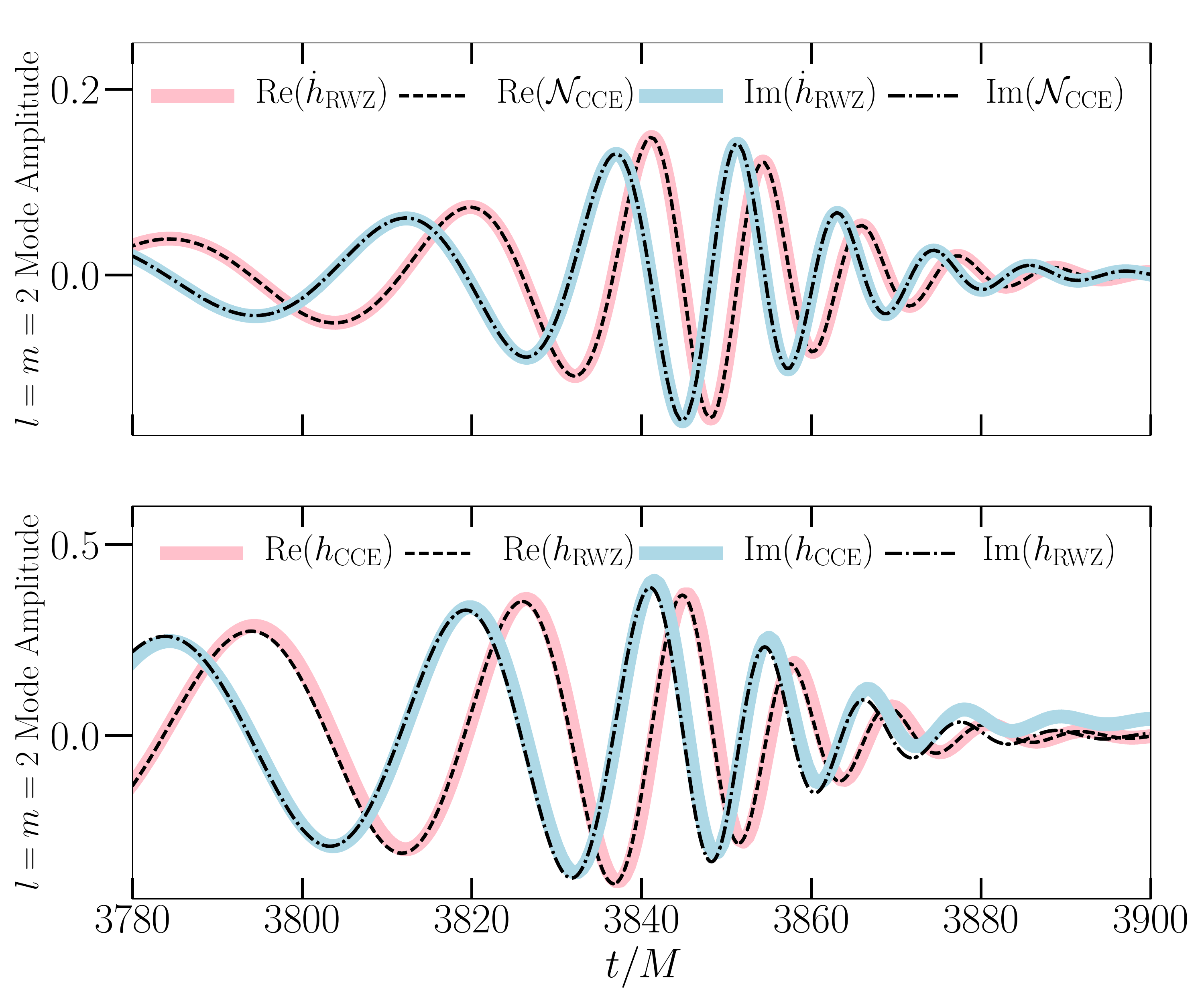} } 
\caption{Comparison between the strain $h$ calculated using CCE and RWZ methods. All waveforms are presented in terms of the $l=m=2$ mode. We use the fact that the strain is the integral of the news function to cross-check the methods. The \textbf{top panel} shows the CCE news function $\mathcal{N}_\mathrm{CCE}$ compared to $\dot{h}_\mathrm{RWZ}$, the derivative of the RWZ strain. The \textbf{bottom panel} shows $h_\mathrm{CCE}$, the integral of the CCE news function, compared to the RWZ strain $h_\mathrm{RWZ}$. We find good agreement until late times, when $h_\mathrm{CCE}$ begins to drift due to the numerical integration scheme (Simpson's rule).}
\label{fig:Strains}
\end{figure}
}

\usepackage{draftwatermark}

\SetWatermarkAngle{90}
\SetWatermarkLightness{1.0}
\SetWatermarkFontSize{0.5cm}
\SetWatermarkScale{1}

\newcommand{\Caltech}{\affiliation{Theoretical Astrophysics,
    Walter Burke Institute for Theoretical Physics,\\
    California Institute of Technology, Pasadena, California 91125, USA}}
\newcommand{\Cornell}{\affiliation{Center for Radiophysics and Space
    Research, Cornell University, Ithaca, New York 14853, USA}}
\newcommand{\Syracuse}{\affiliation{Syracuse University, Syracuse, NY 13244, USA}}

\begin{document}

\title{On choosing the start time of binary black hole ringdown}

\author{Swetha Bhagwat}
\email{spbhagwa@syr.edu}
\Syracuse
\author{Maria Okounkova}
\email{mokounko@tapir.caltech.edu}
\Caltech
\author{Stefan W.~Ballmer}
\Syracuse
\author{Duncan A.~Brown}
\Syracuse
\author{Matthew Giesler}
\Caltech
\author{Mark A.~Scheel}
\Caltech
\author{Saul A.~Teukolsky}
\Caltech
\Cornell


\begin{abstract}

The final stage of a binary black hole merger is ringdown, in which the
system is described by a Kerr black hole with quasinormal mode perturbations.
It is far from straightforward to identify the time at which the ringdown
begins. Yet determining this time is important for precision tests of the
general theory of relativity that compare an observed signal with quasinormal
mode descriptions of the ringdown, such as tests of the no-hair theorem. We
present an algorithmic method to analyze the choice of ringdown start time in
the observed waveform. This method is based on determining how close the strong
field is to a Kerr black hole (\textit{Kerrness}). Using numerical relativity
simulations, we characterize the Kerrness of the strong-field region close to
the black hole using a set of local, gauge-invariant geometric and algebraic
conditions that measure local isometry to Kerr. We produce a map that
associates each time in the gravitational waveform with a value of each of
these Kerrness measures; this map is produced by following outgoing null
characteristics from the strong and near-field regions to the wave zone. We
perform this analysis on a numerical relativity simulation with parameters
consistent with GW150914- the first gravitational wave detection. We find that
the choice of ringdown start time of $3\,\mathrm{ms}$ after merger used in the
GW150914 study~\cite{TheLIGOScientific:2016src} to test general relativity corresponds to a high dimensionless
perturbation amplitude of $ \sim 7.5 \times 10^{-3}$ in the strong-field
region. This suggests that in higher signal-to-noise detections, one would need
to start analyzing the signal at a later time for studies that depend on the
validity of black hole perturbation theory.

    
-
\end{abstract}

\maketitle
 


\section{Introduction}

The quasi-normal mode (QNM) spectrum seen during the ringdown of a perturbed black hole (BH) is determined by the Teukolsky equation; it carries the signature of the BH potential along with the BH horizon and asymptotic boundary conditions~\cite{Teukolsky1, Teukolsky2, Teukolsky3}. The recent detections of binary black hole (BBH) gravitational wave (GW) signals by LIGO (the Laser Interferometer Gravitational-Wave Observatory)~\cite{Abbott:2016blz, Abbott:2016nmj, TheLIGOScientific:2016pea, Abbott:2017vtc, PhysRevLett.119.141101} allow us to begin to probe this QNM signature~\cite{TheLIGOScientific:2016src}. The QNM spectrum in a gravitational-wave observation allows us to perform tests of the no-hair theorem. This theorem states that vacuum, asymptotically flat, stationary, axisymmetric, uncharged BHs are completely characterized by two parameters: the mass and the spin~\cite{Mazur:2000pn,misner1973gravitation,Dreyer,Gossan,Kamaretsos}. This allows us to constrain modified theories of gravity that violate the no-hair theorem~\cite{Berti:2015itd, Yunes:2016jcc}. Observing the QNM spectrum in GWs can be used to validate the BH uniqueness theorem. This theorem states that the exterior geometry of an vacuum, asymptotically flat, stationary, axisymmetric, uncharged BH must be Kerr~\cite{Mazur:2000pn, Mars:1999yn}. 



However, testing the no-hair and uniqueness theorems relies on observing GWs from the QNM perturbative regime (without additional transients remaining from the inspiral). This requires an appropriate choice of start time of this regime.\footnote{While conventions in the literature vary, in this paper, by `ringdown', we explicitly mean the part of the post-merger gravitational waveform that can be described in terms of QNMs.} Identifying this time in the signal is mathematically an ill-defined problem, since QNMs form an incomplete and non-orthogonal basis~\cite{QNMincomplete,QNMnonorthogonal}. Hence, the conventions for choosing the start time of the ringdown have varied in the literature. Berti et al.~\cite{EMOP} and Baibhav et al.~\cite{Baibhav:2017jhs} chose the start time based on maximizing the energy contained in the QNM. London et al.~\cite{London} used $10\,M$ after the peak of the dominant mode of $\Psi_4$ (the Newman-Penrose scalar that encodes outgoing radiation) for fitting to NR waveforms.\footnote{ \label{note1} Since vacuum GR is a scale-invariant theory, it is convenient to express distance and time in terms of source mass by setting $G=c=1$. Explicitly, $ 1~M = M_{\mathrm{BH}} \times G/c^{3} ~\mathrm{seconds}$, where $M_{\mathrm{BH}}$ is the mass of the BH.} Kamaretsos et al.~\cite{2012PhRvL.109n1102K} chose $10\,M$ after the peak luminosity of the dominant mode of the waveform, while Thrane et al.~\cite{Thrane} proposed a loudness-dependent start time. In the  GW150914 testing general relativity (GR) paper~\cite{TheLIGOScientific:2016src}, different start times were used to perform the QNM analysis shown in Fig.~5 of that paper, and the results were consistent with GR when the start time was picked as $3~\mathrm{ms}$ (or later) after the merger.

None of these methods use information from the strong field to motivate the start times. The strong field refers to the region near the BHs (typically within a radius of few $M$), where the scale of the curvature is much smaller than the wavelength of a gravitational wave. In this paper, we develop an algorithmic method for validating choices of the start time of ringdown using strong-field features. Specifically, we measure the \textit{Kerrness}, or closeness to Kerr, in the strong-field region of an NR simulation ringdown, and use null characteristics to map Kerrness onto the GW at asymptotic future null infinity, $\fni$. We then demonstrate this method on a GW150914-like system. 


Determining Kerrness in the strong-field regime is non-trivial, since one needs a coordinate-invariant way of identifying a metric as Kerr. Necessary and sufficient conditions for a gauge-invariant characterization of local isometry to a Kerr manifold were proposed by Garc\'{i}a-Parrado G\'{o}mez-Lobo in~\cite{lobo16}.\footnote{Throughout this text, \textit{isometry} refers to the smooth mapping of manifolds equipped with metrics.} We use this set of algebraic and geometric conditions to provide a numerical measure of Kerrness. Previous studies have used multipole moments of the BH apparent horizon~\cite{Owen:2009sb}, horizon spin measurement comparisons~\cite{Scheel:2008rj}, or Petrov classification~\cite{Baker,Campanelli:2008dv,Owen:2010vw} to characterize ringdown spacetimes. Our work is the first set of conditions that completely characterizes a spacetime as isometric to a Kerr manifold. We study the violation of these conditions post-merger in the strong field of a BBH simulation.

Connecting the strong-field region to the wave zone is a challenge, as the simulation gauge is different from the gauge in which GWs are observed. There is no straightforward way to transform between these gauges. Furthermore, establishing simultaneity between events is not possible in the GR framework, and thus there is no direct map between an event in the strong-field region and a point on the waveform. We therefore devise a scheme to approximately associate the two frames. The association used in this study is of a cause-effect nature: we follow the outgoing null characteristics from the strong-field region to the wave zone using a Cauchy Characteristic Extraction scheme (CCE)~\cite{Handmer:2014qha, Handmer:2015dsa, Casey}, and associate events in the strong field to the wave zone. However, given that GR is a nonlinear theory, the source associated with a particular feature in the GW signal may not be well localized in the spacetime. Nevertheless, one would expect that the source dynamics that dominantly contribute to certain features in the waveform be localizable to a certain extent. Several such approximate localizations have been performed in linear perturbation theory~\cite{PriceAndPlunge,VitorsECO}. 

This paper is organized as follows. Sec.~\ref{sec:Theory} presents the theoretical methods used in this paper, and Sec.~\ref{sec:Implementation} discusses their implementation in NR simulations. Sec.~\ref{sec:ResultMain} then presents and discusses the results of applying these methods to an NR simulation with GW150914-like parameters. We conclude in Sec.~\ref{sec:Conclusion}. Figs.~\ref{fig:PercentOnTheWave} and~\ref{fig:TGR} are the flagship figures, presenting our major results. The the results are quantitatively summarized in Table~\ref{tab:combined info}.

\subsubsection*{Conventions}
We work with the standard 3+1 decomposition of NR (cf.~\cite{baumgarteShapiroBook} for an introduction). In this paper, $g_{ab}$ refers to the spacetime metric, $n^a$ refers to the timelike unit normal vector, $\gamma_{ij}$ refers to the spatial metric on each slice, $D_i$ is the covariant derivative with respect to $\gamma_{ij}$, and $K_{ij}$ refers to the extrinsic curvature. We set $G = c = 1$ and express all quantities in terms of $M$, the sum of the Christodoulou Masses of the two BHs at the start of the simulation. Latin letters at the start of the alphabet, $\{a, b, c, d\}$, refer to (4-dimensional) spacetime indices, while Latin letters in the middle of the alphabet, $\{i,j,k,l,m,n\}$ are (3-dimensional) spatial indices. We denote complex conjugation by an overbar (e.g. $\bar{A}$). To avoid confusion among the multiple meanings of `data' in this paper, we refer to the vacuum data $\{\gamma_{ij}, K_{ij}\}$ on a spatial slice simply as `a slice'.\footnote{\textit{Vacuum data} means that the spatial metric, $\gamma_{ij}$, and the extrinsic curvature $K_{ij}$ satisfy a set of constraint equations corresponding to the decomposition of the vacuum Einstein equations.} Similarly, rather than being purely geometric, a `slicing' in our case is a foliation equipped with a coordinate chart.


\section{Theory}
\label{sec:Theory}
\subsection{Characterizing strong-field Kerrness}

\FlowChartFigure

First, we explain our method of measuring Kerrness in the strong-field region and develop a method to map it onto  $\fni$. Secs.~\ref{sec:KerrnessBackground}~and~\ref{sec:KerrnessTheory} discuss theoretically characterizing Kerrness in the strong-field region, while Secs.~\ref{sec:AFQMotivation},~\ref{sec:Challenges},~and~\ref{sec:AFQ_theory} discuss mapping strong-field information onto the wave zone via null characteristics. 

\subsubsection{Overview and historical background}
\label{sec:KerrnessBackground}
Our overall goal in this section is to evaluate \emph{Kerrness}: how close a numerical BH ringdown spacetime is to being locally isometric to the Kerr spacetime. In order to evaluate the Kerrness of a spacetime, we first need a set of theoretical conditions to evaluate whether a spacetime \textit{is} isometric to Kerr. We can then turn these conditions into a set of \textit{measures}, where deviation from zero indicates being farther from being locally isometric to Kerr. In a numerical simulation, one would evaluate these measures on spatial slices of a simulation. To characterize Kerrness in the strong-field region, one needs local quantifiers evaluated close to the BH, as opposed to looking at regions far away which are contaminated by gravitational radiation. Consequently, we seek a point-wise measure and do not use global measures on a slice such as those proposed in~\cite{Backdahl:2010cy,Backdahl:2010fa,Backdahl:2011np}.

Uniquely characterizing a spacetime as Kerr has been historically challenging---until recently one could only classify spacetimes up to a Petrov type, which gives a weaker classification that admits several manifolds besides Kerr. The Petrov classification uses algebraic properties of the Weyl tensor $C_{abcd}$ based on the four principal null directions (PNDs), by solving the eigenbivector problem (cf.~\cite{stephani2009exact} for a review)
\begin{align}
\label{eq:eigenvalue}
\frac{1}{2} C^{ab}{}_{cd} X^{cd} = \lambda X^{ab}\,,
\end{align}
where eigenbivectors $X^{ab}_{(\alpha)}$ have eigenvalues $\lambda_{(\alpha)}$. The degeneracies of the PNDs give a unique algebraic classification of a spacetime. A spacetime with no repeated PNDs is fully general (Petrov Type I). A spacetime with at least one repeated PND is \textit{algebraically special}. The Kerr metric belongs to a particular class of algebraically special spacetimes, the set of type D spacetimes, which have two double PNDs. A necessary condition for the manifold to be locally isometric to Kerr is to be type D.

Campanelli et al.~\cite{Campanelli:2008dv} used this approach to analyze a numerical BBH ringdown. They determined the degeneracies between the PNDs by solving the eigenbivector problem and measuring the difference between eigenvalues. Their analysis found that the spacetime first numerically settled to type II, which has only one double PND, and then to type D. Owen~\cite{Owen:2010vw} later showed that this measure was sensitive to the choice of tetrad used to compute the Weyl scalars needed to solve the characteristic equation. He proposed a new measure, less-sensitive to the choice of tetrad, and showed that the spacetime settled to type D without first settling to type II.

A type D spacetime can then be shown to be locally isometric to Kerr through additional conditions. Kerr belongs to the Kerr-NUT subset of type D spacetimes. One needs to show that a spacetime is Kerr-NUT and then constrain the acceleration and the NUT parameters. We give more information on Kerr-NUT spacetimes and the various parameters in Appendix~\ref{appendix:KerrNUTParameters}. \saul{Ref.}~\cite{Campanelli:2008dv} investigated the asymptotic behavior of the acceleration and the NUT parameter on a BBH simulation and showed they were constrained to be those of Kerr. 

In this study, we do not solve the eigenbivector problem, but rather use a set of local algebraic and geometric conditions recently proposed by Garc\'{i}a-Parrado G\'{o}mez-Lobo~\cite{lobo16} to show that a spacetime is locally isometric to Kerr. These conditions are formulated in a fully covariant way and thus avoid the complications in~\cite{Campanelli:2008dv} and~\cite{Owen:2010vw} due to tetrad choice. 

\subsubsection{Necessary and sufficient Kerrness conditions}
\label{sec:KerrnessTheory}

To characterize a spatial Cauchy slice as isometric to Kerr, we first check if the slice is algebraically special. Next, we use two geometric conditions to check for the existence of Killing vectors (KVs) on the slice, and we impose two algebraic conditions to verify that the slice containing the KVs is type D. Then, we check the properties of the KVs and further classify the slice into the Kerr-NUT subfamily. Finally, imposing conditions on the acceleration and NUT parameters, we completely characterize the slice as locally isometric to Kerr. These conditions are summarized in Fig.~\ref{fig:FlowChart}. 


All algebraic conditions are expressed in terms of electric and magnetic parts of the Weyl tensor, 
\begin{align}
E_{ab} &\equiv + C_{acbd}n^c n^d \,, \\
B_{ab} &\equiv -{}^*C_{acbd}n^c n^d \,,
\end{align}
where the left dual of the Weyl tensor is defined as ${}^* C^{abcd} \equiv \frac{1}{2}\epsilon^{abef}C_{ef}{}^{cd}$. For a vacuum spacetime, these spatial tensors can be more readily evaluated on a slice as
\begin{align}
\label{eq:electricweyl}
E_{ij} &= K_{ij}K^k{}_k - K_{i}{}^k K_{jk} + {}^{(3)}R_{ij} \,, \\
B_{ij} &= -\epsilon_{kl(i}D^k K_{j)}^l \,,
\end{align}
where ${}^{(3)}R_{ij}$ is the spatial Ricci tensor evaluated from $\gamma_{ij}$. These can be combined into a complex quantity as
\begin{align}
\mathcal{E}_{ij} \equiv \frac{1}{2}\left(E_{ij} - iB_{ij}\right) \,.
\end{align}

In~\cite{lobo16}, the algebraic condition for a slice to be locally algebraically special is given in Eq.~85 as 
\begin{align}
\label{eq:Speciality Index} 
\textrm{\textbf{Speciality Index:\;}} 6b^2 - a^3 = 0  \,,
\end{align}
where 
\begin{align}
a &\equiv 16 \Ep_{ij}\Ep^{ij} \,,  \nonumber  \\ 
b &\equiv -64 \Ep_i^k \Ep^{ij} \Ep_{jk}  \nonumber \,. 
\end{align}
This condition is equivalent to the speciality index in the Petrov classification literature (cf. Eq.~4.13 of~\cite{stephani2009exact}). 

Recall that algebraic speciality corresponds to having one double PND, and hence is a weaker condition than being type D, which corresponds to having two double PNDs. A \textit{necessary} algebraic condition for a slice to be type D is given in Theorem~4 of~\cite{lobo16} as
\begin{align}
\label{eq:typed1}
\textrm{\textbf{Type D 1}}&: \dfrac{a}{12} \gamma_{ij} - \dfrac{b}{a} \Ep_{ij} - 4 \Ep_i{}^k \Ep_{jk} = 0 \,,
\end{align}
which makes use of 4-dimensional algebraic conditions proven in~\cite{Ferrando01} and orthogonally splits these onto the spatial slice. Here we have called the condition `Type D  1' purely for bookkeeping purposes, in order to label each of the type D conditions.

The three \textit{sufficient} conditions for a slice to be type D consist of two geometric conditions involving KVs and one algebraic condition which also includes the KV. As proven in Theorem~2 of~\cite{lobo16}, a vacuum type D spacetime has a complex KV field $\xi^a$ which satisfies an algebraic condition 
\begin{align}
\label{eq:Xi}
\Xi_{ab} = \dfrac{27}{2} w^{\frac{11}{3}} \xi_a \xi_b \,,
\end{align}
where $\Xi_{ab}$ is derived from the Weyl tensor, and 
\begin{align}
w &\equiv -\dfrac{b}{2a} \,.
\end{align}

However, one must show that a KV field exists on the slice in the first place, and then that it satisfies the properties given in Eq.~\eqref{eq:Xi}. The necessary and sufficient geometric conditions for a slice to contain a KV field are known as Killing Initial Data (KID), and for a vector $\xi^a = Yn^a + Y^a$, are given as 
\begin{align}
\label{eq:typed3}
\textrm{\textbf{Type D 3}}&: D_{(i} Y_{j)} - Y K_{ij} = 0 \,, \\ 
\label{eq:typed4}
\textrm{\textbf{Type D 4}}&: D_i D_j Y - \mathcal L_{Y^l} K_{ij} \\
& - Y({}^{(3)}R_{ij} + K K_{ij} - 2 K_{il} K^l_j) = 0 \,. \nonumber  
\end{align}  
Satisfying these conditions guarantees that a KV field exists on the slice---note that these two conditions say nothing so far about type D. 

We can then relate this KV field $\xi^a$ to the condition on the KV in a type D spacetime given in Eq.~\eqref{eq:Xi} by requiring
\begin{align}
\label{eq:typed2}
\textrm{\textbf{Type D 2}}&: \Ep_{pj}(\Omega^2 + \Omega_l \Omega^l) \,, \\
&- 2  \Omega^l\left ( i \Ep_{(p}^k \varepsilon_{j)lk} \Omega + \Ep_{l(p} \Omega_{j)}\right) \nonumber \\
& + \gamma_{pj} \left(\frac{1}{2} w \Omega^2 + \Omega^l \left(-\frac{1}{2} w \Omega_l + \Ep_{lk} \Omega^k  \right) \right) \nonumber \\
& + \frac{1}{2} w \Omega_p \Omega_j  - \dfrac{27}{2} w^{11/3} Y_p Y_j = 0 \nonumber \,, 
\end{align}
where Eq.~\eqref{eq:typed2} is the orthogonal splitting of Eq.~\eqref{eq:Xi}, and 
\begin{align}
\label{eq:definitions}
\Omega_j &\equiv D_k w \,, \\
\Omega &\equiv K^{jk} \Ep_{jk} - w K - 16 i \dfrac{w}{a} \Ep^{jk} \varepsilon_{kpl} D^l \Ep_j^p \nonumber \,, \\
Y &\equiv (w \Omega_j \Omega^j + 2 \Ep_{jk} \Omega^j \Omega^k)^{1/2} w^{-11/6} \nonumber \,, \\
Y_j &\equiv \dfrac{\Omega(2 \Ep_{jk} \Omega^k + w \Omega_j) - 2i \varepsilon_{jkl}\Ep_p{}^l\Omega^p\Omega^k }{27 Y w^{11/3}} \nonumber \,.
\end{align}
This procedure is shown in Theorem 6 of~\cite{lobo16}.\footnote{The Type D 2 condition has a $+$ in the second term where~\cite{lobo16}  has a $-$. The sign error has been confirmed by the author of~\cite{lobo16}. Similarly, The factor of $\frac{1}{27}$ in the definition of $Y_j$ is not included in~\cite{lobo16}, but is in the corresponding Mathematica notebook~\cite{loboprivate}.} 

Type D 3 and Type D 4 are independent geometric conditions that depend on the complex KV $\xi^{a}$ and show that the slice is KID. Type D 1 is a purely algebraic condition that informs us of the behavior of the PNDs. Type D 2 ties in the algebraic and geometric conditions, thereby completing the classification into type D. Speciality Index, meanwhile, is an independent algebraic condition. 

In order to then show that an algebraically special, type D slice is locally isometric to Kerr, we must also show that it belongs to the Kerr-NUT subset of type D spacetimes. Kerr-NUT spacetimes have the symmetry property of two commuting KVs~\cite{stephani2009exact} - one spacelike and timelike, and thus if we impose this geometric condition on KV $\xi^a$ as defined above, we arrive at the condition given in Theorem 8 of~\cite{lobo16},\footnote{However, this has a typographical error (confirmed by the author~\cite{loboprivate}), and should include $\bar Y_j$, the complex conjugate, as given Eq.~\eqref{eq:Kerr1}.}
\begin{align}
\textrm{\textbf{Kerr 1}}&: \operatorname{Im}(Y \bar Y_j) = 0\,.
\label{eq:Kerr1}
\end{align}

In order to further show that a slice is locally isometric to Kerr, we must place constraints on the parameters characterizing Kerr-NUT spacetimes. We summarize the parameters involved in Type D spacetimes in Appendix~\ref{appendix:KerrNUTParameters}. We require that $\lambda$, the NUT parameter, vanish, and $\epsilon$, which is related to the acceleration of the BH, be greater than zero. These conditions are given in Theorem 8 of~\cite{lobo16} as
\begin{align}
\textrm{\textbf{Kerr 2}}&: Z^3 \bar{w}^8 \in \mathbb{R}^{-} \,,
\label{eq:Kerr2}
\end{align}
for the condition $\lambda = 0$, where $Z \equiv \nabla_a w \nabla^a w$, and 
\begin{align}
\textrm{\textbf{Kerr 3}}&: -|Z|^2 + 18\mathrm{Re}(w^3 \bar{Z}) > 0 \,,
\label{eq:Kerr3}
\end{align}
for $\epsilon > 0$. However, the above expression only holds outside of the ergoregion~\cite{loboprivate} in Kerr. This condition is thus impractical to use in the this study, since it involves finding the ergoregion, and masking this region would introduce high levels of numerical error within a spectral code.

 
Thus, for a slice to be locally isometric to Kerr, it must satisfy all of the above conditions, which are summarized in Fig.~\ref{fig:FlowChart}. Since the vacuum spacetime at the start of a ringdown may be fully general, the left hand sides of the Kerrness conditions will not necessarily be zero on some slices. Instead, the Kerrness conditions turn into a set of \textit{Kerrness measures}, where larger deviation from zero indicates a larger deviation from being isometric to Kerr. 

\subsection{Connecting strong-field information to $\fni$}
\subsubsection{Motivation}
\label{sec:AFQMotivation}

\cartoon

Having characterized the Kerrness in the strong-field region, we connect this information to the GWs at $\fni$. We develop a framework to map the evolution of the Kerrness measures computed during a post-merger simulation to the evolution of the post-merger waveform in the asymptotic frame. This provides a procedure to validate the choices of start time of ringdown when analyzing a gravitational-wave signal. 

Just after the two BHs merge, the newly formed BH is expected to be highly distorted. The dynamics of the BH can be explained only via a full numerical simulation. At $\fni$, where the GWs are observed, these strong-field dynamics are responsible for features in a small region close to the peak of the GW amplitude. Once the excitation amplitude in the strong-field region decays to a level when linear perturbation theory is valid 
the spacetime dynamics and the associated waveform is governed by the Teukolsky equation~\cite{Teukolsky1,Teukolsky2,Teukolsky3}. At $\fni$, the waveform appears as a sum of exponentially damped sinusoids with a specific QNM frequency spectrum (with power-law tails that are usually very weak). Beyond this rough picture, the association of the specifics in the strong-field dynamics to the waveform is not well understood, especially during the merger and post-merger phases. 

Understanding this association is crucial because several strong-field tests of GR rely on BH perturbation theory and thus, on identifying the perturbative regime in the waveform. These tests include the no-hair theorem test, consistency tests of the QNM spectrum with the inspiral parameters, and the area theorem test. The start of ringdown in the GW is mathematically ill-defined as damped sinusoids form an incomplete and non-orthogonal basis~\cite{QNMincomplete,QNMnonorthogonal}. Therefore, it is important that we validate the choices of start times in the data analysis of ringdown guided by the strong-field information, where the validity of perturbation theory can be better understood.


\subsubsection{Conceptual challenges}
\label{sec:Challenges}

Mathematically, GR being a non-linear theory does not allow for unambiguous localization of sources of GWs. However, to a certain extent, one expects that the dominant source of a particular feature in the wave zone be localizable to a relatively small region of the spacetime in the past light cone. For instance, studies like~\cite{PriceAndPlunge,PriceAndPlunge2} identify the dominant source for the peak of the waveform during the plunge of a test particle into a Schwarzschild BH with the particle crossing the light-ring.\footnote{The light-ring is the orbit of a massless particle around the BH, which corresponds to the peak of the BH potential located at $3\,M$ in Boyer-Lindquist coordinates for a Schwarzschild BH.} Furthermore, the last few cycles of the BBH GW signal are associated with the dynamics of a linearly perturbed BH~\cite{Vishveshwara:1970zz}. However, one needs to bear in mind that these studies are performed using linear perturbation theory where such localizations are better defined. For example, if one adds a massive particle instead of a test particle in the former case and makes the problem non-linear, one would get some additional source contributions from self-force, thus making the source localization trickier.  

  


In the case of BBH post-merger, identifying specific events as a source of the features in the waveform cannot be done unambiguously owing to the non-linear dynamics from merger. However, drawing intuition from analytical linear perturbation theory, we expect the region within the support of the analytical effective BH potential to contribute significantly to the waveform at $\fni$. Thus, we argue that even in a non-linear case, a small region in the spacetime around the BH containing the strong-field dynamics, can be associated as a dominant source of features in the GW. 

Another challenge in performing this association is that the notion of simultaneity in GR is not absolute, which means that all spacelike slicings of the spacetime are equally valid. In numerical simulations however, we have to make a gauge choice. In our case this choice is made by the Cauchy evolution code. The spatial features corresponding to a particular timeslice are gauge dependent. We choose to monitor the Kerrness on a spatial coordinate 2-sphere in the strong-field region, instead of computing a volume integral over the source region in a timeslice.\footnote{By doing so, the gauge effect is limited to uncertainty of picking the 2-sphere, thereby avoiding contribution of gauge effects through the entire volume region.}


Further, note that this association is only meant to be approximate, and is dependent on the radius of the 2-sphere we monitor. Roughly speaking, the size of the 2-sphere indicates the error bar in the association that originates from the gauge choice (specifically, error $\propto$ radius of extraction) .  


\subsubsection{Forming a source-effect association via null characteristics}
\label{sec:AFQ_theory}


Given these challenges, we propose the following association scheme. By looking at $\Psi_{4}$ evaluated on the simulation, we infer a 2-sphere radius that lies within the strong-field region, containing and generating significant radiative fields. This 2-sphere acts like an effective source for the GW seen at $\fni$. We evaluate a surface integral of the Kerrness measures at each time slice during the ringdown on this 2-sphere. Then, we connect the evolution of the Kerrness measures on this surface to the associated features in the GW by following the outgoing null characteristics emanating from this 2-sphere. The details of this procedure are described below. 

The GWs emanating from a source propagate to $\fni$ along outgoing null rays (since the spacetime is curved, a small portion of GWs also travel inside the light cone). We utilize this in constructing an association between strong-field information and the features on the GW.  We associate a feature on the GW to a 2-sphere in the strong-field region at a given time (in the simulation coordinates) if they lie on the same outgoing null hypersurface. This 2-sphere can thus be interpreted as an effective source producing the point on the waveform. Hence, the choice of 2-sphere should be close to the region generating GWs rather than farther out. Measuring Kerrness of such a surface would give an insight into validity of perturbation theory in the region that acts as a dominant source of the GWs. 

A framework that is naturally suited for such connections is Cauchy Characteristic Extraction (CCE). CCE foliates the spacetime into a family of outgoing null hypersurfaces and formulates Einstein's equations as an initial-boundary value problem in a 2+2 characteristic decomposition. The mathematical details of this formalism can be found in~\cite{Bishop,Casey}. CCE performs a characteristic evolution using the metric data on a timelike boundary of the Cauchy region (known as the worldtube) and propagates it to $\fni$. At $\fni$ the radiation information is obtained as the Bondi news function  $\mathcal{N}$~\cite{Bondi}. The GW strain can then be obtained from $\mathcal{N}$ by a time integration,

\begin{align}
h(t)= \int_{-\infty}^{t} \mathcal{N}(t') dt'\,.
\end{align} 


 A key feature of this scheme is that each point at $\fni$
 corresponds to a null hypersurface, which in turn corresponds to a particular (coordinate) time label on the world tube.

 We can thus associate the average of the Kerrness on a 2-sphere to spherical harmonic modes at $\fni$. We illustrate this in Fig.~\ref{fig:Cartoon}. Here $\tau_{0}$ marks a specific timeslice  (horizontal solid green line) in the Cauchy evolution region in a gauge chosen by the Cauchy code. The intersection of this timeslice with the worldtube boundary is a spatial (topological) 2-sphere. The information on this 2-sphere is propagated to $\fni$ along a null hypersurface labeled (solid purple line) as $\tau_{0}$. The radiation feature carries the time stamp $\tau_{0}$ at $\fni$, which, roughly speaking, arises from the 2-sphere defined by the intersection of timeslice $\tau_{0}$ and the worldtube in the simulation and thus, we identify them to be associated.  

Having established a framework to associate information on a 2-sphere in the strong-field region to the waveform at $\fni$, we now discuss the choice of the 2-sphere used in this study. Motivated by analytical studies of test particles plunging into Schwarzschild BHs \cite{PriceAndPlunge,PriceAndPlunge2}, one might want to inspect the 2-sphere associated with the peak of effective BH potential. However, locating it during the merger in a numerical simulation is non-trivial (if at all well-defined), and is beyond the scope of this paper. Furthermore, CCE cannot be performed from an arbitrarily small worldtube close to the horizon. This limitation arises because CCE is formulated in light-cone coordinates. In the regions very close to the horizon, light-cone coordinates can form caustics, leading to coordinate singularities. Because of these constraints, we choose the worldtube radius corresponding to the smallest coordinate  2-sphere that is accessible to our procedure, but we visually verify that it contains strong-field dynamics by plotting $\Psi_{4}$ in Figs.~\ref{fig:psi41B}~and~\ref{fig:psi42B}.

\subsection{Inferring perturbation amplitudes via Kerrness}
\KerrPertAmplitudeFigure
\EnvelopesFigure
In order to give physical meaning to the values of the Kerrness measures outlined in Sec.~\ref{sec:KerrnessTheory}, we can compare their values (on a post-merger spacetime, for example) to those on a single BH with a known analytic perturbation. Specifically, we can compare the Kerrness measures during ringdown to those on a $l=m=2$ spheroidal QNM perturbed Kerr BH of the same final mass and spin, with varying dimensionless perturbation amplitude $\varepsilon$. This will provide a true physical comparison, as linearly-perturbed type D spacetimes are fully generic type I, and thus the Kerrness measures on the perturbed spacetime are expected to be nonzero~\cite{Araneda:2015gsa}. This comparison will allow us to infer the perturbation amplitude to which a particular coordinate time corresponds. We can then map this inferred amplitude onto the waveform using the methods in Sec.~\ref{sec:AFQ_theory}.

\subsubsection{Kerrness measures on perturbed metrics}
\label{sec:PerturbationTheory}


The perturbed metric is generated on a single slice for each $\varepsilon$ by solving the Teukolsky equation and reconstructing the metric perturbation $h_{ab}$ using a Hertz-potential formalism~\cite{Yang:2014tla, Lousto:2002em} (cf.~\cite{Teukolsky:2014vca} for a general review). The resulting perturbation $h_{ab}$ is then added to the background metric to give 
\begin{align}
\label{eq:perturbation}
\tilde g_{ab} = g_{ab}^\mathrm{Kerr} + \varepsilon h_{ab}\,,
\end{align}
which is correct to linear order. The constraint equations for the metric $\tilde g_{ab}$ are then solved to give a fully constraint-satisfying metric $g_{ab}$ in Kerr-Schild coordinates using the extended conformal thin-sandwich formalism (cf.~\cite{baumgarteShapiroBook}).
This introduces some nonlinear effects into the perturbed metric. Furthermore, the asymptotic radial behavior leads to blow-up of the solution at large radii~\cite{Ori:2002uv}. Thus, before solving for $g_{ab}$, we multiply $h_{ab}$ by an envelope of the form 
\begin{align}
\label{eq:envelope}
f_\mathrm{Envelope}(R) = \exp[-((R - r_+)/W)^F/2] \,,
\end{align}
where $r_+$ is the radius of the outer horizon of the BH, $W$ is the width, and $F$ is the falloff of the envelope. Since the mapping of the Kerrness measures onto the waveform occurs at  $R=5\,M$, as will be discussed in Sec.~\ref{sec:AFQ_implementation}, and the horizon typically has outer radius $R_+ \sim 1.7\,M$, we choose $W = 6\,M$ to give $f_\mathrm{Envelope}(5\,M) \sim 1$ so as to minimally affect the perturbation at the extraction radius. Additionally, we choose $F = 8$ in order to counteract the steep growth of the perturbation with radius. We plot the envelope in Fig.~\ref{fig:Envelopes}. In practice, the metric perturbation is generated using an extension of the code used in East et al.~\cite{East:2013mfa}, but with the QNM solution rather than an ingoing GW solution and using the full radial dependence. 



Fig.~\ref{fig:KerrPertAmplitude} shows the behavior of the Kerrness measures averaged on a 2-sphere of $R=5\,M$ with $\varepsilon$ on a BH of the same final mass and spin as the simulation outlined in Sec.~\ref{sec:simulation}. The theoretical behavior of the Kerrness measures with perturbation amplitude is unknown~\cite{loboprivate, Ionescu:2014cta}, and thus this is the first (numerical) computation of the behavior. We first check that the measures converge to finite values with numerical resolution, thus representing real physical values. The Kerrness measures increase quadratically for small $\varepsilon$, then show higher-order effects for large $\varepsilon$. Type D 2 grows to (best-fit) quartic, Type D 3 and Kerr 1 become cubic, while Specialty and Type D 1 remain quadratic at $\varepsilon \sim 10^{-2}$, the largest amplitude for which we can solve for $g_{ab}$. In particular, the steep increase of the Type D 3 and Kerr 1 measures, which come from geometric conditions on KVs, indicates that at large enough perturbation amplitude, the slice fails to have even an approximate KV. Since the perturbation we are introducing is not axisymmetric, it makes sense that at large $\varepsilon$ the slice loses this KV symmetry. 

The linear perturbation regime corresponds to the region where the measures increase quadratically with $\varepsilon$, while the non-linear regime approximately begins where one can see higher-power behavior. In this case, we see the transition from quadratic behavior around $\varepsilon_\mathrm{critical} \sim 5 \times 10^{-3}$, suggesting that this is the approximate start of the nonlinear regime. In practice, one can normalize all of the $\varepsilon$ values in this paper by $\varepsilon_\mathrm{critical}$. However, we do not do this for readability of the figures. 

However, there are some sources of error in the $g_{ab}$ analysis. The areal radius of the perturbed metric on a coordinate 2-sphere of radius $R = 5\,M$ changes slightly with perturbation amplitude, changing by $10^{-2}\,M$ between $\varepsilon = 10^{-6}$ and $10^{-2}$. Thus, a coordinate-radius measure comparison does not happen on exactly the same 2-sphere. Solving for $g_{ab}$ changes the values of the mass and spin from the parameters used in creating $g_{ab}^\mathrm{Kerr}$. At the largest perturbation amplitude $\varepsilon = 10^{-2}$, the dimensionless spin changes by $.003$, while the mass changes by $.008\,M$. We keep these errors in mind when computing the Kerrness values of the strong-field region in terms of $\varepsilon$ and mapping them to the waveform for the binary case in Sec.~\ref{sec:PerturbationResults}. 

\subsubsection{Mapping onto the waveform}
\label{AmpMapOnFni}
A perturbation amplitude $\varepsilon$ is associated with each timeslice of a post-merger spacetime in the strong-field region by the procedure described above. Since the procedure developed in ~\ref{sec:AFQ_theory} allows us to associate simulation timeslices with the gravitational waveform at $\fni$, we can map the perturbation amplitude associated with each timeslice to the corresponding parts of the waveform at $\fni$.  This gives an insight into deciding which portion of the waveform at $\fni$ can be modeled as being generated by linearly perturbed Kerr manifold, thus providing validation of start times chosen in data analysis that rely on perturbative description of Kerr. 

\subsection{Measuring Kerrness on the horizon}
\label{sec:MultipolarTheory}

In addition to local measures throughout a spatial slice discussed in Sec.~\ref{sec:KerrnessTheory}, Kerrness can also be evaluated on the post-merger apparent horizon (AH) (also known as the \textit{dynamical horizon} in the literature). Owen describes a multipolar horizon analysis in~\cite{Owen:2009sb}, finding that the multipolar structure of a final BBH remnant was that of Kerr. Probing the multipolar structure also serves as a test of the no-hair theorem~\cite{Teukolsky:2014vca}. 

This formalism involves computing the mass multipole moments $I_\alpha$ of the horizon as 
\begin{align}
I_\alpha = \oint y_\alpha R dA\,,
\end{align}
where $R$ is the scalar curvature of the horizon, $dA$ is the metric volume element on the AH, and $\alpha$ labels generalized (non-axisymmetric) scalar spherical harmonics $y_\alpha$. These generalized spherical harmonics are computed from the eigenvalue problem 
\begin{align}
\Delta y_\alpha = \lambda(\alpha) y_\alpha\,,
\end{align}
where $\Delta$ is the operator $\Delta \equiv g^{AB} \nabla_A \nabla_B$ on the AH, and $\lambda$ is its eigenvalue. In analogy with axisymmetric spherical harmonics $Y_{lm}$, an effective $l$ is defined for the eigenvalues as 
\begin{align}
\label{eq:leff}
\lambda = -\dfrac{l_\mathrm{eff}(l_\mathrm{eff} + 1)}{r^2}\,,
\end{align}
where $r$ is the areal radius of the horizon. Since the $l_\mathrm{eff}$ values are time-dependent, we refer to a given multipole by its final value. 

As discussed in~\cite{Owen:2009sb}, the multipole moments that are zero on a Kerr BH either immediately vanish due to the symmetry of the dynamical horizon, or decay to zero from their excited values as the remnant BH settles to Kerr. The multipole moments that do not vanish on Kerr are functions of the mass and spin, and reach these values with increasing coordinate time. We use the code implemented and tested in~\cite{Owen:2009sb} to compute the multipole moments. However, since the multipole moments are features of the horizon, we cannot map their behavior onto the waveform at $\fni$. Moreover, CCE cannot be performed close to the horizon, as discussed in Sec.~\ref{sec:AFQ_theory}. Nevertheless, we can compare the qualitative behavior of the multipole moments with those of the Kerrness measures as done in Secs.~\ref{sec:MultipolarResults} and~\ref{sec:PercentKerrnessmap}.


\section{Numerical implementation}
\label{sec:Implementation}

\subsection{Description of simulation}
\label{sec:simulation}

We apply the methods outlined Sec.~\ref{sec:Theory} to the numerical simulation presented in Fig.~1 of~\cite{PhysRevLett.116.061102}, with similar parameters to GW150914, the first LIGO detection. The simulation is performed and the methods are applied using SpEC, the Spectral Einstein Code. The waveforms and parameters are available in \texttt{SXS:BBH:0305} in the SXS Public Catalog~\cite{SXS:catalog}. The simulation has initial mass ratio $q = 1.221$, and dimensionless spins $\chi_A = (0, 0, 0.33)$ and $\chi_B = (0, 0, -0.44)$. The initial orbital frequency is $\Omega_0 = 0.017$. The final (post-merger) BH has dimensionless spin $\chi_C \simeq (0, 0, 0.69)$ (within numerical error, as measured using the techniques in~\cite{Scheel:2008rj}) and mass $0.952\,M$. The inspiral proceeds for $3694.4\,M$ until the formation of a fully-resolved common AH. The visible part of the post-merger waveform on a linear scale has a temporal duration of $\sim 61\,M$. 

Within a BBH simulation, the metric equations are evolved in a damped harmonic gauge~\cite{Szilagyi:2009qz, Lindblom2009c}, with excision boundaries just inside the apparent horizons~\cite{Hemberger:2012jz, Scheel2014}, and minimally-reflective, constraint-preserving boundary conditions on the outer boundary~\cite{Rinne2007}. The spectral grid used during the inspiral of the simulation has an excised region for each BH. Once a common AH forms, the simulation proceeds for a few timesteps before switching to a new grid, in which there is one excision region for the new AH~\cite{Hemberger:2012jz}. For this simulation, the grid-switch happens at $3696.9\,M$. For more information on the code, see~\cite{Lovelace:2016uwp}.

\subsection{Implementation of Kerrness measures}
\label{sec:numericalimplementation}

We discuss the numerical implementation of the Kerrness measures outlined in Sec.~\ref{sec:KerrnessTheory}, and summarized in Fig.~\ref{fig:FlowChart}, on an NR BBH post-merger. Note that these measures will not be zero even on a numerical Kerr spacetime, due to the resolution of the simulation. 

\ConvergenceTestFigure


In order to quantify the Kerrness measures at each point, we convert the complex tensors into scalars by contracting them as
\begin{align}
\label{eq:normalization}
S_A = A^{ij}\bar{A}_{ij} \;\;\;\mathrm{and}\;\;\; S_B = B^{i}\bar{B}_i \,,
\end{align}
where raising and lowering occurs using the spatial metric $\gamma_{ij}$.\footnote{The Kerr 2 measure given in Eq.~\eqref{eq:Kerr2} requires that the imaginary part be zero, while the real part be $\geq 0$. Hence, when evaluating Kerr 2, we measure the deviation of the imaginary part from zero, and the deviation of the real part from being positive (hence only including negative values).} Throughout the rest of the paper, all of the measures will refer to their respective scalars generated using Eq.~\eqref{eq:normalization}. 



Because our simulations are performed using spectral methods, we expect errors to converge exponentially with increasing numerical resolution~\cite{Press:2007:NRE:1403886}.  In Fig.~\ref{fig:ConvergenceTest}, we plot the Kerrness measures as a function of resolution for a single Kerr black hole; we see that the measures decay exponentially towards zero as expected.

SpEC solves a first-order formulation of the Einstein equations, and therefore evolves both the spacetime metric and variables corresponding to its time and spatial derivatives~\cite{Lindblom:2005qh}. The metric and first derivatives are available to the accuracy of the numerical simulation on each slice. Kerrness measures that require additional numerical derivatives, however, will have greater numerical noise and a higher numerical noise floor. The highest numerical order derivative needed to evaluate each measure is given in Fig.~\ref{fig:FlowChart}. Type D 4, which requires four numerical derivatives, is the noisiest measure and has a higher noise floor than the other measures\Mark{,} as shown in Fig.~\ref{fig:ConvergenceTest}.

\subsection{Map from source to $\fni$ - implementation}
\label{sec:AFQ_implementation}
\alg

\phiplot

In our study, we use a CCE implementation in SpEC (cf.~\cite{Kevin}, in prep). This implementation uses
a no ingoing and outgoing radiation condition on the initial null hypersurface of the characteristic evolution. This means that the code treats the spacetime outside the worldtube as initially free of any gravitational radiation from the past.\footnote{During the Cauchy evolution, we perform the evolution with a boundary of $R \approx 670\,M$ and we do not neglect the backscatter from the region outside of the CCE extraction radius.} Usually the CCE worldtube is placed at a large radius, and the CCE evolution begins at the start of the numerical simulation during early inspiral. However, here we begin CCE only at the merger portion of the Cauchy evolution, and in addition, we place the CCE worldtube at a very small radius. This means that extracted waveform does not contain contribution coming from the inspiral part of the dynamics. 

By decreasing the radius of the extraction worldtube progressively by $1\,M$, we find the smallest radius of the worldtube that our procedure can be applied to occurs at a coordinate radius of $R= 5\,M$. For a radius of $R=3\,M$, the CCE procedure can not be performed, presumably due to the formation of caustics. At $R=4\,M$, we get a very glitchy and unreliable extraction of the news function. 

However, performing the CCE from such small radii gives rise to an additional complication. Since time stamps on the waveform at $\fni$ are induced by the simulation coordinates, the news function obtained is not necessarily in an inertial gauge. In a standard CCE scheme, a gauge transformation is applied to the news function in order to obtain it in an inertial gauge.  To preserve the map between the time in simulation gauge and the time coordinate on the extracted news function, we do not perform this gauge transformation. We see the effect of the gauge transformation in the waveform at $\fni$ as a mixing of mode amplitudes. The effect is very small when the worldtube boundary for CCE is large i.e., lies in the weak field region. For instance, for a worldtube boundary of $R=128\,M$ the effect of this transformation is negligible. To confirm this, we compute the overlap $\mathcal{O}$ between the news extracted from $R=128\,M$ with and without the gauge transformation. The overlap $\mathcal{O}$ is defined as, 
\begin{align}
\label{eq:overlap}
\mathcal{O} = \left\langle \mathcal{\widetilde{N}}_{1} | \mathcal{\widetilde{N}}_{2}\right\rangle = \int_{- \infty}^{\infty} \frac{\mathcal{\widetilde{N}}_{1}(f) \mathcal{{\widetilde{N}}}_{2}^{*} (f)}{|\mathcal{\widetilde{N}}_{1}||\mathcal{\widetilde{N}}_{2}|}~df\,,
\end{align}
where $\mathcal{\widetilde{N}}_{1,2}$ is the frequency domain Fourier-transformed news function, and ${}^*$ denotes complex conjugation for ease of readability, and $||$ is the norm~\cite{overlap}. 

We find that the mismatch, $1 - \mathcal{O}$, is $ \sim 10^{-6}$. This overlap computation uses only the merger and post-merger parts of the news function for the dominant ($l=m=2$) spin-weighted spherical mode. However, for a worldtube radius of $R=5\,M$, there could be significant amplitude deviations between the waveforms in the simulation-coordinate-induced gauge and the inertial gauge. Because of technical difficulties in the code implementation, we could not apply the gauge transformation to an extraction from $R=5 M$ and quantify the difference. \footnote{In the CCE implementation we use, the inertial coordinate at $\fni$ is setup by solving and quantity $\omega$ used in definition of the conformal factor in~\cite{Casey, caseyprivate}. The possible technical issues when performing extraction from small radius are - a) initializing the inertial coordinates are tied to the Cauchy evolution coordinates and b) the $\omega$ evolution scheme may not be accurate in the strong field regime. The interpolation scheme is not designed for extraction performed from very small radii.} 

Furthermore, before the non-inertial to inertial gauge transformation, every point on $\fni$ at the same timestamp on the waveform corresponds to the same null hypersurface and therefore to the same simulation coordinate time. After the transformation, this is no longer true: the waveform seen at different sky directions with the same timestamp on the waveform corresponds to different null hypersurfaces and therefore different values of simulation coordinate time. This happens because the choice of the 2-sphere is gauge-dependent.Therefore, we omit the gauge transformation, as the aim in this paper is to connect the near-zone to the wave zone, requiring us to retain the timestamps.


Additionally, the initial no-ingoing radiation condition neglects gravitational radiation coming from the inspiral. This may be significant for extraction done at small radii, where the initial CCE null hypersurface connects the strong-field region close to merger to $\fni$ and may contain significant radiation from the inspiral. This could contribute towards the discrepancy between the $R=128\,M$ and $R=5\,M$ waveforms. 


To assess this difference, we compare the news function obtained by extraction performed from $R=5\,M$ with the extractions performed from the worldtubes of larger radii, all without the gauge transformation. The result of this is presented in Figure \ref{fig:AlignmentAtMaximumOftheNews}. We observe that all the extractions from radii greater than $32\,M$ converge with radius, indicating that the effect of the gauge transformation is insignificant at these radii. Further, the extraction from $R=5\,M$ has a significant amplitude discrepancy with the other extractions, particularly in its first cycle. Therefore, we would ideally wish to map the strong-field information computed on the 2-sphere at a coordinate radius of $R=5\,M$ on the news function that has been extracted from a larger radius like $R=128\,M$. 



We do this mapping in two steps. First, we map the strong-field information computed on the 2-sphere at a coordinate radius of $R=5\,M$  onto the CCE performed from a worldtube of  $R=5\,M$ using the framework described above.  Next, we note that the phase evolution of extraction from $R=5\,M$ agrees with the extractions from larger radii.\footnote{The time-derivative of the phase gives the instantaneous frequency of the gravitational radiation.} We verify this in Fig.~\ref{fig:phasediffcomb}. Then we align the news function extracted from $R=5\,M$ to the extraction from larger radii as shown in  Fig.~\ref{fig:AlignmentAtMaximumOftheNews}. The alignment is done such that the overlap $\mathcal{O}$ between the CCE extracted news function from different world tube radii is maximized. The maximum normalized $\mathcal{O}$ between the news function extracted from $R=128\,M$ and $R=5\,M$ is $0.82$. Incidentally, this alignment is equivalent to aligning the real part of the news function at its global minima (or global maxima of the absolute value). Table \ref{tab:Tshift} lists the time shifts that have been applied in order to align the news function extracted from a radius $R_{i}$ with extraction done at $R=128\,M$. 
\begin{table}[!htb]
    \centering
        \begin{tabular}{ r@{ = }l@{}l| r@{.}l@{}l} \hline \hline
\multicolumn{3}{c |}{Worldtube radius} &\multicolumn{3}{l}{Alignment shift wrt $R=128\,M$}\\ \hline
$R$&$5$&$\,M$ & $132$ & $5$&$\,M$ \\
$R$&$32$&$\,M$ & $96$ & $5$&$\,M$ \\
$R$&$64$&$\,M$ & $62$ & $5$&$\,M$\\
$R$&$128$&$\,M$ & $0$ & &$\,M$ \\
\hline \hline
\end{tabular}
        \caption{The shift in the time axis performed to align the news functions extracted from different radii in Fig.~\ref{fig:AlignmentAtMaximumOftheNews}. The alignment has been done such that the overlap between the news function extracted from different worldtube radii with the extraction from $R=128\,M$ is maximized.
        }
\label{tab:Tshift}
\end{table}

Using this alignment we map the time stamps on the $R = 5\,M$ to those on $R= 128\,M$. From this, we infer the mapping of strong-field information at $R=5\,M$ on to the extraction done from $R=128\,M$, thus mapping the strong-field information onto the news function as seen in near inertial gauge. 

We summarize our algorithm for mapping the strong-field information onto the news function: 
\begin{enumerate}
\item Perform CCE from worldtube with radius of the 2-sphere that lies in the strong-field region (whose evolution you wish to map on to the news function seen at $\fni$) without the final non-inertial to inertial gauge transformation. The time stamps on this extracted news function are induced by the time coordinates in the simulation, thus providing a natural map between the evolution of the strong-field region and the wave zone.
\item Perform CCE from a large worldtube radius where the effect of the non-inertial to inertial gauge transformation is negligible. 
\item Align the news functions obtained in steps 1 and 2 such that the overlap between the waveform is maximized. 
\item Use this alignment to map the time stamps of the news function extracted in step 1 to that in step 2. The 2-sphere chosen in step 1 at the timeslice marked with the simulation time coordinate can be associated as the dominant source of the feature at $\fni$ with the same time stamp. 
\end{enumerate}

\section{Results}
\label{sec:ResultMain}

We now present the results of performing the analysis outlined in Secs.~\ref{sec:Theory}~and~\ref{sec:Implementation} on the GW150914-like simulation detailed in Sec.~\ref{sec:simulation}. Sec.~\ref{sec:MultipolarResults} presents the behavior of the multipole moments of the AH, which provides a comparison for the Kerrness measures on the simulation volume. Sec.~\ref{sec:PercentKerrnessmap} discusses the results of evaluating the Kerrness measures on the post-merger spacetime and mapping them onto the waveform at $\fni$, presenting them in terms of the percentage decrease from their peak values. Sec.~\ref{sec:PerturbationResults} presents the results of comparing the Kerrness measures on the post-merger spacetime to values on perturbed data, in order to infer the perturbation amplitude in the strong-field region, and mapping them onto the waveform, presenting them in terms of the inferred perturbation amplitude $\varepsilon$. The percentage decrease from the peak value and $\varepsilon$ can then be used to estimate the overall level of Kerrness and validate choices for the start time of ringdown. Finally, in Sec.~\ref{sec:DataAnalysis}, we discuss the implications of these results on analyzing ringdown in GW data, and in Sec.~\ref{sec:TestingGRComparison} we compare our results to the ringdown start times chosen in the GW150914 testing GR study~\cite{TheLIGOScientific:2016src}.

\subsection{Horizon behavior and multipolar analysis on BBH ringdown}
\label{sec:MultipolarResults}

\HorizonDataFigure

As a first measure of Kerrness, we apply the horizon multipolar analysis outlined in~\cite{Owen:2009sb} and summarized in Sec.~\ref{sec:MultipolarTheory} to the simulation described in Sec.~\ref{sec:simulation}. Fig.~\ref{fig:HorizonData} presents the behavior of the AH. The areal mass of the AH, given by $\sqrt{A/16\pi}$ where $A$ is the proper area of the AH, sharply settles to a final value. The minimum and maximum radii are initially noisy, as the AH is initially peanut shaped, but they decrease exponentially with coordinate time, showing a settling of the AH to the final state. However, the radii are coordinate-dependent measures, and thus to check if the BH settles to Kerr it is more instructive to look at the AH multipole moments.

Fig.~\ref{fig:HorizonData} shows the behavior of the initially non-vanishing quadrupole and hexadecupole moments, labeled by their corresponding $l_\mathrm{eff}$ at the final time, as given in Eq.~\eqref{eq:leff}. The quadrupole moments correspond to $l_\mathrm{eff} \sim 2$ and the hexadecapole moments correspond to $l_\mathrm{eff}\sim 4$. The multipole moments behave as expected for a generic simulation remnant settling to a Kerr BH. As explained in~\cite{Owen:2009sb}, two of the five quadrupole moments immediately vanish by reflection symmetry, while two others exponentially go to zero (eventually hitting a numerical noise floor) as the final remnant settles to Kerr. Four of the nine possible hexadecupole moments immediately vanish from reflection symmetry, while four go exponentially to zero as the remnant settles to Kerr. Note that the $l = 1$ and $l = 3$ moments vanish on Kerr due to symmetry. As in~\cite{Owen:2009sb}, one quadrupole moment ($l_\mathrm{eff} = 2.1$) and one hexadecupole moment ($l_\mathrm{eff} = 4.17$), both corresponding to $m = 0$, do not vanish, but rather attain a constant value in line with that of a Kerr BH of the same final mass and spin.

The multipolar behavior thus confirms that the final state of the AH is that of a Kerr BH. This serves as an independent test of Kerrness, and thus one would expect the Kerrness measures presented in Sec.~\ref{sec:KerrnessTheory} to also show the strong-field region exponentially settling to Kerr. This also serves as numerical evidence for BH uniqueness, as the final remnant of a BBH merger is indeed Kerr, as also discussed in~\cite{Owen:2009sb}. Similarly, since the final multipolar structure can be described completely by the mass and spin, this serves as numerical validation of the no-hair theorem. 

\subsection{Measuring and mapping Kerrness onto the waveform}
\label{sec:PercentKerrnessmap}

 The goal in this section is to validate choices of the start time of ringdown using Kerrness measures on the GW150914-like system described in Sec.~\ref{sec:simulation}. We now present the results of evaluating the Kerrness measures outlined in Secs.~\ref{sec:KerrnessTheory} and~\ref{sec:numericalimplementation} (and summarized in Fig.~\ref{fig:FlowChart}) in the strong-field region and mapping them onto the waveform at $\fni$ using the procedure given in ~\ref{sec:AFQ_implementation}. These measures are evaluated point-wise on each slice,  and we map the value on a 2-sphere at a radius of $R= 5\,M$ onto the news function. Recall that larger values of the Kerrness measures indicate greater deviation from being locally isometric to Kerr. 
 
 \RainbowFigure

Fig.~\ref{fig:Rainbow} shows the Kerrness measures averaged at various coordinate radii on each slice of the post-merger spacetime, presented as a function of coordinate time. All of the measures decay exponentially toward zero, showing that the spacetime approaches an isometry to Kerr. This confirms the results of the multipolar analysis in Sec.~\ref{sec:MultipolarResults}. Additionally, this serves as a numerical verification of BH uniqueness, as the final state of a BBH merger is isometric to Kerr. The behavior of the measures at large radii (such as $R=128\,M$ in this case) is especially interesting to the question of BH uniqueness, which is particularly concerned with the \textit{domain of outer communication}~\cite{Ionescu:2014cta}. 

Fig.~\ref{fig:Swirl} shows the behavior of the Speciality Index, an algebraic measure (Type D 1) and a geometric measure (Kerr 1) in the volume, as a function of increasing coordinate time. We see a distinct quadrupolar pattern in all our measures (the equatorial plane has a modal pattern that corresponds to $|m| = 2$), consistent with the dominant mode of gravitational radiation. Furthermore, the Speciality Index  and Type D 1 measures, which determine properties of the PNDs, settle first further from the BH, while the geometric Kerr 1 measure, which is determined by properties of the KV, first settles closer to the BH. 

\SwirlFigure

The Kerr 2 measure, which constrains the NUT parameter, is effectively constant throughout the ringdown, as shown in Fig.~\ref{fig:KerrTwo}. Since the NUT parameter is one of the hairs of a generic type D manifold, Fig.~\ref{fig:KerrTwo} confirms that a NUT charge is not generated during a BBH merger. We thus do not include it further in our analysis. 

\KerrTwoFigure

Of these measures, two are algebraic constraints---Type D 1 and Type D 2---and three are geometric constraints on the KV, Type D 3, Type D 4, and Kerr 1. In Fig.~\ref{fig:Rainbow} we see that all the algebraic measures decay in a similar fashion and all the geometric measures decay similarly. Type D 4, which requires 4 numerical derivatives, is visibly noisier than the other measures. This measure checks if the vector identified as $(Y,Y_{j})$ satisfies the Killing equation and is crucial for a rigorous mathematical characterization of Kerr manifold. However, all geometric measures depend on the same Killing vector and we observe that Type D 4 has a similar decay property as Type D 3 and Kerr 1. Thus, we do not include the noisier Type D 4 in our analysis, rather treating Type D 3 as a proxy for both. 



Each measure at each radius in Fig.~\ref{fig:Rainbow} eventually reaches a floor. This is confirmed to be a numerical noise floor in Fig.~\ref{fig:NoiseFloor}, where the floor is shown to exponentially converge to zero with numerical resolution. The radial behavior of the Kerrness measures stems from the radial behavior of the Weyl tensor and the metric quantities. For example, for a stationary background, $E_{ij} \sim R^{-3}$ and $B_{ij} \sim R^{-4}$, and thus Speciality Index given in Eq.~\eqref{eq:Speciality Index} should be $\sim R^{-18}$, which we indeed observe. 

\NoiseFloorFigure

The analysis outlined in Sec.~\ref{sec:AFQ_implementation} requires the Kerrness measures to be extracted at $R=5\,M$ in order to map them to the news function. Fig.~\ref{fig:Swirl} shows that the Kerrness measures have strong support at $R=5\,M$, thus justifying the choice of radius as being in the near field.\footnote{The measures at $R=3\,M$ in Fig.~\ref{fig:Rainbow} behave similarly to those at $R=5\,M$ indicating that $R=3\,M$ also behaves like the near field region, but unfortunately  we have not been able to perform CCE from this small a radius.}

The Kerrness measures quantify the violation of the conditions for a manifold to be isometric to Kerr and therefore, they need not have the same dimensions and sensitivities. Thus, one cannot compare the absolute magnitudes of these measures with each other and directly translate their value into statements on validity of start time of perturbative regime. In order to normalize and combine them into an overall measure of Kerrness,  we use the concomitant percentage decrease from their peak values. 

We present the percentage decrease of each of these measures from their peak values mapped on to the news function in Fig.~\ref{fig:PostItPanel} and Fig.~\ref{fig:specialityIndex}. In the bottom panels of these figures, the news function is plotted as a function of time. On the same time axis, the top panel depicts the corresponding evolution of the Kerrness measure in the strong-field region. The waveform feature in the bottom panel at a particular time coordinate is associated to the timeslice carrying the same time label, via source-effect association outlined in Sec.~\ref{sec:AFQ_theory}. In the bottom panel, the Kerrness value at this time characterizes the deviation from Kerr. 
\postIt

\postItSpecial

In these figures, we delineate 6 lines marking the percentage decrease from the peak value of each of the Kerrness measures as a function of time---both in the strong-field region and on the news function at $\fni$. As stated before, these measures have different decay properties and so do not decay to a particular percentage of their peak value at the same time. The difference between the time at which measures decay to a particular percent is tabulated in Table~\ref{tab:percentageDecrease}.  

\begin{table}[!htb]
    \centering
            \begin{tabular}{ r@{ }l |  r@{.}l@{}l | r@{.}l@{}l} \hline \hline
\multicolumn{2}{c |}{$\%$ of peak value} & \multicolumn{3}{c |}{Spread in time} & \multicolumn{3}{l}{Combined \% time} \\ \hline
100 & $\%$ & 12& &$\,M$ & 3847&5 &$\,M$ \\
50 & $\%$ & 9&8 &$\,M$ & 3857& &$\,M$ \\
30 & $\%$ & 9& &$\,M$ & 3860&7 &$\,M$ \\
10 & $\%$ & 8&3 &$\,M$ & 3867&7 &$\,M$ \\
5 & $\%$ & 8&7 &$\,M$ & 3871&9 &$\,M$ \\
1 & $\%$ & 6&1 &$\,M$ & 3881&3 &$\,M$ \\
\hline \hline
\end{tabular}
    \caption{The spread in the time for given $\%$ of the peak value of Kerrness measures computed using all the measures. The combined $\%$ time refers to the value of the dashed lines in  Fig.~\ref{fig:PercentOnTheWave} and corresponds to the time at which all the measures have at least decayed to the indicated $\%$. }
\label{tab:percentageDecrease}
\end{table}

We present the combined percentage decrease from the peak value on the news function in Fig.~\ref{fig:PercentOnTheWave}. The shaded bands correspond to spread in percentage decay on the news function. The widths of these bands are given in Table~\ref{tab:percentageDecrease}. The solid line at the end of each band marks the time when all these measures have decayed to the indicated percentages and this can be used to conservatively choose the start time. 

Furthermore, in this figure we do not include the Specialty Index. The Specialty Index is an independent measure that quantifies if the manifold is algebraically special. Since this is the weakest condition in our Kerrness characterization scheme, we see that it gets satisfied earliest on the post-merger simulation from Fig.~\ref{fig:specialityIndex}. The $1 \%$ of peak line which occurs unexpectedly late arises because of numerical reasons. We assert this by looking at the nearly flat nature of Specialty Index curves in Fig.~\ref{fig:Rainbow} at late times, very close to the numerical noise floor. 

\perOnWave

We observe that all measures decay to $\sim 50 \%$ of their peak value within half a cycle from the peak of the news function. Further, in approximately one cycle, all the measures are reduced to $\sim 30 \%$ of their peak values. The spread in each of the bands is about $\sim 10\,M$ when we include all the Kerrness measures in computing the band, and this shrinks to $\sim 6\,M$ when we exclude Specialty Index. 

We combine the measures with equal weights, thereby presenting a conservative result. Furthermore, we have repeated our analysis with larger worldtube radii and confirmed that our results do not change significantly; specifically, for $R =~32,~64$ and $80\,M$ the results change only by $\sim 4\,M$. 



\subsection{Estimating and mapping the perturbation amplitude onto the waveform}
\label{sec:PerturbationResults}

\PsiFourComparisonFigure

In order to provide a physical understanding for the values of the measures in the strong-field region shown in Figs.~\ref{fig:Rainbow} and~\ref{fig:Swirl}, we can compare the values to those on an initial slice of a perturbed Kerr BH with the same final mass and spin as the BBH simulation, as outlined in Sec.~\ref{sec:PerturbationTheory}. We can then map the inferred strong-field perturbation amplitude $\varepsilon$ onto the waveform using the procedure outlined in Secs.~\ref{sec:AFQ_theory}~and~\ref{sec:AFQ_implementation}. This procedure involves the following steps:

\begin{enumerate}
\item Generate perturbed Kerr manifolds for a range of amplitudes $\varepsilon$.
\item Compute the Kerrness measures at $R=5\,M$ on these slices.
\item Compute the Kerrness measures at $R=5\,M$ on the post-merger BBH simulation (verifying that the gauge-invariant areal radii of the $R=5\,M$ coordinate 2-spheres are approximately equal for the single-BH and the BBH case).  
\item Identify the coordinate time in the post-merger BBH simulation at which the Kerrness measures at $R=5\,M$ agree with those on the perturbed Kerr slice for a given $\varepsilon$ --- this gives a \textit{crossing time} for this $\varepsilon$. 
\item Use this crossing time to map the inferred $\varepsilon$ onto the waveform. 
\end{enumerate}

\CrossingTimeFigure

Fig.~\ref{fig:CrossingTime} shows the inferred $\varepsilon$ for the BBH ringdown simulation as a function of coordinate time in the simulation. The gauge-invariant areal radii at $R = 5\,M$ on the BBH simulation slices and on the metric perturbation are within $10^{-2}\,M$. The values of the Kerrness measures on the perturbed data vary quadratically with $\varepsilon$, as shown in Fig.~\ref{fig:KerrPertAmplitude}. At higher values of $\varepsilon$, they obtain higher-power dependence, as discussed in Sec.~\ref{sec:PerturbationTheory}. Each Kerrness measure decays through various $\varepsilon$ as the simulation progresses. Type D 1 and Type D 2, the two algebraic conditions, have comparable crossing times for a given $\varepsilon$, while the two geometric KV conditions, Type D 3 and Kerr 1, also have comparable crossing times. Speciality Index crosses around $10\,M$ before the other measures, in part because it is a weaker condition that the others. Each crossing time has an intrinsic $2\,M$ spread due to sampling, and not all measures cross each $\varepsilon$ due to numerical noise floors, leading to spreads in crossing time.

\ampmap

In Fig.~\ref{fig:Psi4Comparison}, we visually check the spacetime features by comparing $\Psi_4$ corresponding to $\varepsilon = 7.5 \times 10^{-3}$~and~$10^{-3}$ on the perturbed Kerr metric with the corresponding timeslice during the post-merger simulation. The crossing time spread for a particular $\varepsilon$ arises because of the imperfect mapping between an analytically perturbed Kerr BH and the post-merger spacetime. Therefore, unlike in an ideal mapping, the combined crossing times will have a spread. In particular, the difference in the features between the post-merger and the perturbed Kerr slice indicates a difference in symmetry and explains the larger spread in the crossing time between the  KV-dependent measures. We see that the spread in the combined crossing times using only algebraic measures is much smaller than when we include the geometric measures.

We next map the inferred perturbation amplitude to the news function, using a procedure similar to the one in the previous section, and present the result in Fig.~\ref{fig:crossingtime}. The top panel of the figure indicates the crossing time for the Specialty Index, the middle panel for the algebraic measures, and the bottom panel shows that for geometric measures. The spread in the crossing time for the algebraic measures decreases from $ \sim 6\,M$ at the start, to our sampling rate, $2\,M$. This occurs because at the very start of post-merger, the system is not yet in a perturbative regime and therefore, our mapping contains a larger error. Geometric measures are more drastically affected by the imperfections in the mapping, indicating the differences in the symmetries of the two systems. On including the geometric measures, the crossing time spreads to $ \sim 8\,M$. We confirm that the spread of the crossing times calculated using the algebraic measures is always contained within the spread of crossing times calculated using the geometric measures. 

As the signal decays from the peak to a barely visible amplitude on a linear scale ($\sim 3-4$ cycles) at $\fni$, the corresponding perturbation in the strong-field region decreases by an order of magnitude. The peak of the news function corresponds to a perturbation amplitude of $\sim 7.5 \times 10^{-3}$. Further, it takes about 2 cycles in the wave zone for the perturbation amplitude to decay to half its peak value. Also, by the time the perturbation amplitude decays by an order of magnitude, there is hardly any power left in the signal. 

\subsection{Implication of the start time on data analysis}
\label{sec:DataAnalysis}
\subsubsection{From news to h}
\label{sec:hoft}
In order to compare the Kerrness measures on the GW to the loss in signal-to-noise ratio (SNR) at the times used in~\cite{TheLIGOScientific:2016src}, we must first calculate the strain $h$ from the news function, and then calculate the merger time. As outlined in Sec.~\ref{sec:AFQ_implementation}, $h$ can be calculated by integrating the CCE news function. One can also independently calculate $h$ using the Regge-Wheeler-Zerilli (RWZ) (cf.~\cite{RZW} for details on the method) ~\cite{PhysRev.108.1063, PhysRevLett.24.737, PhysRevD.2.2141, Moncrief:1974am} method  and then extrapolating it in powers of the extraction radius (cf.~\cite{Boyle:2009vi} for details). The RWZ method and extrapolation  have been implemented and tested in SpEC~\cite{Boyle:2009vi, Taylor:2013zia}, and the strain calculated by this method was presented in the GW150914 detection paper~\cite{PhysRevLett.116.061102}. This method, however, has a different retarded time axis~\cite{Boyle:2009vi} than the CCE news function. Thus, we differentiate the RWZ strain to get a news function, and shift it to align in phase with the CCE news function. We check the CCE results by comparing the output of the two methods, presenting the results in Fig.~\ref{fig:Strains}.

In the GW150914 testing GR study~\cite{TheLIGOScientific:2016src}, $t_\mathrm{merger}$ is defined as the point at which the quadrature sum of the $h_\times$ and $h_+$ polarizations of the most-probable, or \textit{maximum a posteriori} (MAP) waveform, produced by Effective-One-Body (SEOBNRv4) template~\cite{Purrer:2015tud} is maximal. For this study, we use the $l=m=2$ spin-weighted spherical harmonic mode of the MAP waveform, as this is the least-damped QNM. In this study, rather than using the EOBNR waveform, we calculate $t_\mathrm{merger}$ based on the time of maximum amplitude of the time-shifted RWZ strain, as
\begin{align}
\label{eq:tmerger}
t_\mathrm{merger} \equiv \{t | h^2(t) = \max_{t'} (h^2(t'))\} \,,
\end{align}
where 
\begin{align}
h^2 \equiv \mathrm{Real} (h)^2 + \mathrm{Imag} (h)^2\,.
\end{align}
We find $t_\mathrm{merger} = 3839.0 \pm 0.1\,M$. 

\StrainsFigure

\subsubsection{Start time and the SNR}

\begin{table}[!htb]
    \centering
        \begin{tabular}{ r@{ }l | c | c | l} \hline\hline
\multicolumn{2}{c |}{No. of cycles} & $\%$ SNR & $\%$ Kerrness & $\varepsilon/10^{-3}$ \\ \hline
&Peak  & 60 & 100 &  7.5\\
$\frac{1}{2}$ &cycle & 30 - 40 & 40 - 50 &  7.5 \\
1 &cycle & 20 - 25 & 35 - 30 &  5\\
1 $\frac{1}{2}$ &cycles  & 10 - 20 & 8 - 12 &  3.5\\
2 &cycles  &  $\sim$ 10 & 7 - 5 &  2 - 2.5 \\
2 $\frac{1}{2}$ &cycles  &  $<$ 10  & $\sim$ 1 &  1-2 \\
3 &cycles  & $<$  5 & $ <$ 1 &  0.5 - 1 \\  \hline\hline
\end{tabular}
    \caption{Summary of our results. The first column counts the number of cycles from the peak of the news function. The second column presents the drop in SNR with start time chosen in the data analysis. SNR is normalized to have $100 \%$ when the data analysis starts at the peak of the waveform ($h(t)$) i.e., at $3839\,M$. The third column shows the concomitant percentage decrease in the Kerrness measures from the peak value (similar to Fig.~\ref{fig:PercentOnTheWave}). Further, in the last column we present the perturbation amplitude inferred by the crossing times computed with Type D 1 and D 2 measures (similar to middle panel of Fig.~\ref{fig:crossingtime}.) }
\label{tab:combined info}
\end{table}

\SNRdp
\spread

While picking too early a start time for an analysis that relies on being in ringdown gives inaccurate and biased results, picking a start time too late leads to a large statistical error. Since the amplitude of the signal decays exponentially with time, the SNR in ringdown decreases as exponential-squared with the start time. Consequently, the spread in the posteriors during estimation of ringdown parameters, which goes inversely proportional to match-filtered SNR, increases and gives rise to large statistical uncertainties. Therefore, one must chose an optimal middle ground considering both these factors. 

In the top panel of Fig.~\ref{fig:PercentageSNR}, we show the percentage decrease in match-filtered SNR with the start time of the ringdown. A match-filtered SNR is a noise-weighted scalar product between the signal and the template and is defined as
\begin{align}
\mathrm{SNR} = 4 \int_{0}^{\infty} \frac{\tilde{h_{1}^{*}}(f')\tilde{h}_{2}(f')}{S_{h}(f')} df' = \langle h_1|h_2\rangle \,,
\label{eq:SNR}
\end{align}
where ${}^*$ denotes complex conjugation for ease of readability. Here, $h_{1}(t)$ corresponds to a ringdown waveform that is tapered at $t_\mathrm{merger}$ and acts as a signal. We filter this against the template, $h_{2}(t)$, which is tapered with varying start time. Further, $S_h(f)$ corresponds to power spectral density (PSD) of aLIGO at design sensitivity~\cite{newnoise}. However, since we present our results in terms of ratios, our analysis remains valid to any detector noise curve. Then, a Fourier transform is taken to evaluate Eq.~\eqref{eq:SNR}. Here we use only the $l=m=2$ spin-weighted spherical harmonic mode of the RWZ strain waveform computed in Sec.~\ref{sec:hoft}. The system is considered to be optimally oriented with respect to the detector for this calculation.  

The tapering is done with a tanh window function defined as 
\begin{align}
\mathfrak{W} (t) = \mathrm{tanh}[\alpha_{0} (t-t_{0})]/2\,.
\label{eq:tapering}
\end{align}
$t_{0}$ is the time around which the tapering is centered and it is set to the start time of the perturbative regime. $\alpha_{0}$ is set to 10 in making the top panel of Fig.~\ref{fig:PercentageSNR}. Furthermore, we confirm that our results do not change significantly with the tuning parameter $\alpha_{0}$ using $\alpha_{0}= \{2,5,10,20\}\,M^{-1}$.

We then present percentage decrease of SNR in the top panel of Fig.~\ref{fig:PercentageSNR} by defining $100 \%$ for start time at $t_\mathrm{merger}$. Further, on this same plot we also indicate the  amplitude of perturbation in the strong-field region (as calculated using the algebraic measures) at the start time, giving an insight into how statistical and systematic errors vary with the choice of start time. 

The bottom panel of Fig.~\ref{fig:PercentageSNR} presents the total energy radiated through the merger-ringdown as a function of time. This indicates the strength of GW signal and is calculated by integrating~\cite{Ruiz:2007yx}
\begin{align}
\label{eq:EnergyRad}
\dfrac{dE}{dt} = \lim_{r \to \infty} \frac{r^2}{16 \pi} \oint \left| {\int_{-\infty}^t \Psi_4 dt'}\right|^2 d \Omega \,.
\end{align}
Furthermore, on the same plot we mark the percentage decrease of the Kerrness measures from their peak values, providing a comparison between the strength of the signal left for performing the analysis versus Kerrness evaluated in the strong-field region. 

To impress the sharp trade-off in systematic and statistical uncertainties in choosing the start time of the ringdown and, to provide an intuition of implication of Fig.~\ref{fig:PercentageSNR}, we present the spread in estimation of dominant QNM frequency,  $f_{22}$ in Fig.~\ref{fig:spread}. For this, we calculate the spread using the Fisher information matrix formalism similar to that in Eq.~4.1a of~\cite{bertiparam}, for a GW150914-like system. In particular, we set $f_{22}$ to 253.7 Hz and the quality factor, $Q_{22}$ to 3.2. However, we emphasize that this is a rough estimate intended only to provide intuition and, we plan to follow this up by a rigorous Bayesian parameter estimation in the future.


We present the interplay between the systematic and statistical uncertainty concisely in Table~\ref{tab:combined info}. 
Furthermore, we find that by the time the news function peaks, the SNR has already dropped down to $60 \%$. However, at this time the algebraic Kerrness measures are at their peak value. We also observe that by about a cycle of news function, there is less than 20 percent SNR left in the signal. Therefore, there seems to be a sharp trade-off between the systematic modeling error and statistical uncertainties.

\subsection{Comparison with GW150914 testing GR paper }
\label{sec:TestingGRComparison}

\tgr

The test of consistency of ringdown of the GW150914 signal with the analytically predicted QNM frequency is given in~Fig.~5 of~\cite{TheLIGOScientific:2016src}. The analysis chooses various start times of ringdown, namely $t_\mathrm{merger} + 0,~1,~3,~5,~6.5\,\mathrm{ms}$. At a start time of $t_\mathrm{merger} + 3\,\mathrm{ms}$ (or later), parameter estimation of the dominant QNM in ringdown is consistent with predictions using initial masses and spins. 

A time $3\,\mathrm{ms}$ for the system corresponds to $9.4\,M$ from $t_\mathrm{merger}$. In our analysis, $t_\mathrm{merger} = 3839\,M$ (cf. Eq.~\eqref{eq:tmerger}), while the peak of the news function occurs at $3846\,M$. Thus, $3\,\mathrm{ms}$ corresponds to $2.4\,M$ after the peak of the news function. In this region, as shown in Fig.~\ref{fig:TGR}, the perturbation amplitude is ${}\gtrsim 7.5 \times 10^{-3}$. Our analysis indicates that this corresponds to a relatively large deviation from Kerr. Recall that Fig.~\ref{fig:KerrPertAmplitude} suggests that $\varepsilon = 5 \times 10^{-3}$ is the approximate start of the nonlinear regime. 

With a start time of $t_\mathrm{merger} + 3\,\mathrm{ms}$, the SNR was about $8.5$ and the spread in the estimate of QNM frequency was roughly $40\,\mathrm{Hz}$~\cite{TheLIGOScientific:2016src}. Because of this low SNR and high spread, the GW150914 testing GR analysis may not have been sensitive to the large non-Kerrness we see close to the BH. However, in the case of higher SNR, where the analysis is sensitive to the systematics of the ringdown model, our study suggests picking a later start time. 

Our analysis uses geometric and algebraic conditions to identify isometry to Kerr. However, these conditions do not directly measure the deviation of the curvature BH potential from that of Kerr. Since the QNM are intrinsic tests of the BH potential along with the boundary conditions, deviation of QNM frequencies will depend on details of the BH potential, and thus are not directly quantified in our measures. Additionally, the parameters used in this study correspond to \texttt{SXS:BBH:0305} waveform used in the GW150914 detection paper~\cite{Abbott:2016blz}, which are slightly different from those of the MAP waveform used in the testing GR paper.

\section{Conclusion}
\label{sec:Conclusion}



In this study, we present a method for validating choices of
the time at which a BBH GW signal can be considered to enter the ringdown stage. This is done by computing algebraic and geometric measures of Kerrness in the strong-field region of an NR simulation of a BBH ringdown, and then associating each point on the asymptotic-frame waveform with a particular value of these Kerrness measures. Thus, for each point on the asymptotic-frame waveform there is an estimate for how close the BH spacetime is to Kerr spacetime. This is the first time this set of measures, proposed in~\cite{lobo16}, are evaluated in the strong-field region. This is also the first time measures of Kerrness in the strong-field region is mapped onto the waveform.  We outline this method in Secs.~\ref{sec:Theory}~and~\ref{sec:Implementation}, and implement this analysis in Sec.~\ref{sec:ResultMain} on a GW150914-like NR simulation. 

We observe that after merger, the Kerrness measures of a BBH ringdown simulation decrease exponentially with coordinate time in the simulation, eventually settling to a numerical noise floor, as shown in Fig.~\ref{fig:Rainbow}. This decay is consistent with measuring Kerrness using multipole moments of the apparent horizon, as in Fig.~\ref{fig:HorizonData} and~\cite{Owen:2009sb}. In all cases, the measures on the final slice of the simulation indicate that the final remnant is a Kerr BH, thus providing numerical consistency with the BH uniqueness theorem. Moreover, we find that the final state in the multipolar analysis depends just on mass and spin, which serves as a confirmation of the no-hair theorem in the strong-field region. Additionally, as shown in Fig.~\ref{fig:Swirl}, the Kerrness measures have a quadrupolar (with $|m| = 2$) structure consistent with the dominant gravitational radiation. The geometric measures, which rely on the existence of a Killing vector field, first decay to zero close to the horizon, then later they decay at larger radii as gravitational radiation propagates out. On the other hand, algebraic measures, which depend on principal null directions, first decay to zero at larger radii, before decaying near the BH. We also see that  the NUT parameter remains zero during merger and ringdown,
as shown in Fig.~\ref{fig:KerrTwo}. 


These gauge-independent Kerrness measures are crucial to the nonlinear stability analysis of Kerr, as they quantify the deviation from being isometric to Kerr. The analytical behavior of these measures with perturbation amplitude is unknown~\cite{Ionescu:2014cta, loboprivate}. Through this study we provide insights into their numerical behavior in Fig.~\ref{fig:KerrPertAmplitude}. We find that all of these measures scale quadratically with $\varepsilon$ for low amplitude perturbations, but acquire  higher-order nonlinearities for larger perturbation amplitudes. Furthermore, in Figs.~\ref{fig:Rainbow}~and~\ref{fig:Swirl}, we provide the radial behavior of these measures, up to a large radius of $R=128\,M$. For a BBH simulation, we track these measures starting from merger, where linear perturbation theory is not expected to hold. Despite the large initial excitation, our study shows that the BBH ringdown simulation evolves to a final Kerr state, providing a numerical validation of the nonlinear stability of Kerr. 

To connect the Kerrness measures in the strong-field region to the asymptotic waveform at $\fni$, we use CCE, which evolves Einstein's equations on a foliation of outgoing null hypersurfaces.
A null characteristic evolution can be done only in a region free from caustics. We demonstrate that CCE results using a worldtube at $R=5\,M$ are consistent with those done from larger radii. This implies that during ringdown, caustics only exist very close to the BH. Furthermore, we show that the map between the strong-field region and the wave zone can be extended all the way in to $R=5\,M$.



Although caustics do not form, we see in Figs.~\ref{fig:Swirl},~\ref{fig:psi41B}, and~\ref{fig:psi42B} strong features in the curvature quantity $\Psi_4$ in the region enclosed by $R \sim 10\,M$. This implies that our extraction radius choice of $R=5\,M$ lies within the strong-field and within the support of the BH potential.


In Fig.~\ref{fig:PostItPanel}, we label each point of the BBH ringdown waveform with the percentage decrease of the Kerrness measures in the strong-field region relative to their maximum values.
In order to give a physical interpretation of the values of the Kerrness measures, we compare them throughout the post-merger spacetime to those evaluated on a $l=m=2$ QNM perturbed Kerr BH of the same final mass and spin. From this we infer the amplitude of BH perturbation during ringdown and map onto a particular point in the BBH ringdown waveform; this is marked on the BBH ringdown waveform in Fig.~\ref{fig:crossingtime}. 

As the BBH simulation proceeds after merger, the strong-field region starts looking like Kerr, indicating the validity of perturbative analysis. However, as time progresses, the amplitude of the ringdown decreases, leading to a rapid decay in SNR in a GW detection. We find that by the time the Kerrness measures decrease to $50\%$ of their peak values, there is only about $20\%$ SNR left in the signal. In terms of perturbation amplitude close to the BH, this maps to an amplitude between $7.5 - 5 \times 10^{-3}$. This occurs after $1 - 1.5$ cycles of the news function, which corresponds to $\sim 16.4\,M$ after $t_\mathrm{merger}$. Additionally, we find that the
 start time of ringdown used in~\cite{TheLIGOScientific:2016src}, $t_\mathrm{merger} + 3\,\mathrm{ms}$, corresponds to an amplitude of $7.5 \times 10^{-3}$. In future detections with higher SNR, where the statistical noise is significantly smaller, one may need to choose a later start time to perform precision tests of GR such as no-hair theorem tests. 

A future extension to this project would be to investigate methods that allow us to perform similar source-asymptotic frame associations for smaller radii. For instance, the light ring would be an interesting region to monitor as it is crucial to the QNM structure. This can perhaps be done numerically through ray-tracing methods such as those used in~\cite{Bohn:2016afc} and~\cite{Bohn:2014xxa}, to understand the evolution of the peak of the BH potential (if it forms).  Another possible technique could be to try performing CCE from smaller radii after the high amplitude of the initial excitation has reduced. Additionally, being able to perform this association at smaller radii would allow one to understand the propagation of perturbations very close to the BH horizon onto the waveform; these are expected to be redshifted and appear on the waveform with a large time delay. 

Another avenue of future work would be investigating the effects of implementing a more realistic condition on the initial null hypersurface by relaxing the no-ingoing-waves condition used in performing CCE.  In addition, we can study the trade-off involved in choosing an earlier ringdown time, which will decrease the spread in recovered ringdown parameter posterior distributions and increase the systematic errors that arise because of deviations from Kerr in the strong-field region. 

The methods used in this paper can be applied to future BBH detections in order to guide the choice of the start time of ringdown. For the sake of quick reference to the procedure described in this paper, we concisely enumerated the steps in Appendix~\ref{Appedix:Steps}. Note that the results of this paper approximately holds for any equal mass system with an appropriate mass rescaling (cf. footnote~\ref{note1}). Our method would better allow one to perform precision tests of GR that depend on being in perturbative regime, such as tests of the no-hair theorem and area theorem. With this procedure, we provide an algorithmic way to check whether an unexpected deviation in a QNM analysis is due to not being in the perturbative regime, rather than due to a violation of GR or corresponding theorems. 
For future detections, we plan to repeat this analysis using an NR simulation with the MAP waveform parameters.

\acknowledgments

We would like to thank
Alfonso Garc\'{i}a-Parrado G\'{o}mez-Lobo,
Kevin Barkett, 
Mike Boyle,
Yanbei Chen,
Joshua Goldberg,
Casey Handmer,
Daniel Hemberger,
Maximilliano Isi,
Badri Krishnan, 
Nicholas Meyer,
Harald Pfeiffer,
Leo Stein and Vijay Varma
for many valuable conversations. 
In particular, we would like to thank Mike Boyle, Casey Handmer, Harald Pfeiffer, and Leo Stein for careful reading of this manuscript. We would like to thank William East for
helping to generate the QNM metrics, and Geoffrey Lovelace
for supplying the BBH simulation data used in this
study. 
This work was supported in part by the Sherman Fairchild Foundation,
the Brinson Foundation,
NSF grants PHY--1404569 and AST--1333520 at Caltech, and NSF grant PHY--1606654 at Cornell University and PHY--1404395, PHY-1707954 and PHY-1352511 at Syracuse University. 
MO gratefully acknowledges the support of the Dominic Orr Graduate
Fellowship at Caltech.
We used SpEC (Spectral Einstein Code) to perform the simulations and analysis~\cite{spec}.
Computations were performed on the 
Zwicky and Wheeler clusters at Caltech, which are supported by the Sherman 
Fairchild Foundation and by NSF award PHY--0960291. The BBH simulation was performed on  the ORCA cluster at California State University, Fullerton (CSUF), supported by
CSUF, NSF grant No. PHY--142987, and the Research
Corporation for Science Advancement.

\appendix{}

\section{Quick outline of our procedure}
\label{Appedix:Steps}
In this appendix, we concisely provide an outline of the algorithmic procedure  developed in this paper. For future BBH detections, we propose to follow the steps enumerated below- 
\begin{enumerate}
\item Performing an NR simulation with waveform parameters inferred from parameter estimation, and saving the metric data,
\item Generating worldtube data for various extraction radii and performing CCE from the inner-most possible radius,
\item Evaluating the Kerrness measures on the metric data at this radius for BBH ringdown, 
\item Evaluating the Kerrness measures on QNM perturbed data with the same final mass and spin, and inferring corresponding perturbation amplitude from the Kerrness values,
\item Mapping the Kerrness measures and inferred perturbation amplitudes to the waveform via null-characteristics,
\item Using these results to validate choices for the start time of ringdown in detector data analysis.
\end{enumerate}

\section{Kerr-NUT parameters}
\label{appendix:KerrNUTParameters}

In this appendix, we provide a review of the parameters of the Kerr-NUT solution. The Kerr family of vacuum solutions is unique when one imposes axisymmetry, stationarity and regularity on the BH horizon along with asymptotic flatness. However, if one allows for generalization by relaxing the asymptotic flatness condition, one arrives at a family of solutions called Kerr-NUT. This solution is a part of the broader family of Einstein-Maxwell type D solutions. This generalized family of spacetimes is parameterized by 6 parameters (potentially 7 if one includes the cosmological constant $\Lambda$). In Table~\ref{tab:Parameters}, we summarize the parameters, as well as their physical meaning and symbols used in various texts.  

The general Einstein-Maxwell Type D solution (including cosmological constant $\Lambda$) has the form given in Eq.~21.11 of~\cite{stephani2009exact}, with parameters $m$, $l$, $\gamma$, $\varepsilon$, $e$, and $g$. $m$ refers to the mass parameter (closely related to the mass of the BH), $\gamma$ is related to the angular momentum parameter $a$ (closely related to the spin of the BH), $\varepsilon$ is related to the acceleration $b$, $e$ is the electric charge, $g$ is the magnetic charge, and $l$ is known as the NUT parameter. As outlined in~\cite{PLEBANSKI197698}, the mass and the NUT parameter form a complex quantity, as do the angular momentum and the acceleration, similarly to the electric and magnetic charges. In~\cite{PLEBANSKI197698}, $\varepsilon$ and $\gamma$ do not appear in the curvature quantities, and are called kinematical parameters, while the others are dynamical parameters. 

As shown in Table~21.1 of~\cite{stephani2009exact}, setting all of the parameters to zero except for $m$, $a$ (and hence $\gamma$ and $\varepsilon$), and $e$ to zero yields the Kerr-Newman solution, while also setting $a = 0$ yields the Reissner-Nordstrom solution. Kerr-Taub-NUT metrics, meanwhile, are parametrized by mass, spin, and $l$, with $l \neq 0$, and are thought to be unphysical~\cite{kaluzaklein}. The vacuum BBH case considered in this study, meanwhile, sets $e = 0$ and $g = 0$, since there are no electric or magnetic charges at the start of the simulation, and no sourcing of them during the simulation. 

An accelerating and rotating BH with a NUT charge will have non-zero $m$, $l$, $a$, and $b$, with $a > l$. A Kerr solution with a NUT charge will then have $b = 0$. An accelerating and rotating BH, meanwhile, will have $l = 0$. Finally, the Kerr solution has both $l = 0$ and $b = 0$. An illustration of this is provided in Fig.~1 of~\cite{Griffiths:2005se}. The condition $l = 0$ gives the Kerr 2 condition considered in this paper, given in Eq.~\eqref{eq:Kerr2}.

After setting $l=0$, the parameters $m$, $\varepsilon$ and $\gamma$ are related to the mass and spin of a BH are as follows, 
\begin{align}
\label{eq:massAndspin}
\mathrm{mass} = \frac{m}{\varepsilon^{\frac{3}{2}}} \; \; \; \; \mathrm{and} \; \; \; \;
\mathrm{spin} = \frac{2 \sqrt{|\gamma|}}{\varepsilon} \,.
\end{align}

Since, $\varepsilon > 0$ and $m>0$ for a Kerr BH, the condition that $b = 0$ gives $\varepsilon > 0$, which corresponds to the Kerr 3 condition given in Eq.~\eqref{eq:Kerr3}.

\newcommand*\rot{\rotatebox{90}}

\begin{table}[h]
  \begin{tabular}{ l | c  c  c  c}
     & \rot{Stephani~\cite{stephani2009exact}} & \rot{Garc\'{i}a-Parrado~\cite{lobo16}} & \rot{Plebanski~\cite{PLEBANSKI197698}} & \rot{Griffiths~\cite{Griffiths:2005se}}\\
    \hline \hline
    Cosmological constant & $\Lambda$ & & $\lambda$ & \\ \hline
Mass parameter & $m$ & $\mu$ & $m$ & $m$  \\ \hline
NUT parameter & $l$ & $\lambda$ & $n$ & $n$ \\ \hline
Angular momentum parameter& $\gamma$ & $\gamma$ & $\gamma$ & $k$  \\ \hline
Acceleration parameter& $\varepsilon$ & $ \epsilon$ & $\epsilon$ & $\epsilon$  \\ \hline
Electric charge & $e$ & & $e$ & $e$ \\ \hline
Magnetic charge & $g$ & & $g$ & $g$
  \end{tabular}
\caption{%
	Parameters of the family of the Einstein-Maxwell type D solutions, presented with physical meanings in the rows and naming conventions in various literature in the columns. These parameters do not measure the physical quantities directly but are intimately connected to the physical quantities they describe. For instance, Eq.~\eqref{eq:massAndspin} shows how the mass and spin of a BH are related to the mass parameter and the angular momentum parameter. 
 }
  \label{tab:Parameters}
\end{table}

\bibliography{Biblo}
\end{document}